\documentclass[aps,pre,showpacs,preprintnumbers,twocolumn,nofootinbib]{revtex4-1}
\usepackage{geometry}                
\usepackage{graphicx}
\usepackage{amssymb}
\usepackage{dsfont}
\usepackage{epstopdf}
\usepackage{color}
\usepackage{mathtools}
\usepackage{hyperref}

\definecolor{red}{rgb}{1,0,0}
\definecolor{blue}{rgb}{0,0,1}
\definecolor{black}{rgb}{0,0,0}
\newcommand{\blue}{}

\newcommand{\p}{\partial}
\newcommand{\eq}[1]{\begin{align}#1\end{align}}
\newcommand{\eqs}[1]{\begin{align*}#1\end{align*}}
\newcommand{\ffrac}[2]{\mbox{$\frac{#1}{#2}$}}
\newcommand{\half}{\mbox{$\frac{1}{2}$}}
\newcommand{\OO}{\mathcal{O}}
\newcommand{\tr}{\mbox{tr}}

\usepackage{scalerel,stackengine}
\stackMath
\newcommand\widecheck[1]{%
\savestack{\tmpbox}{\stretchto{%
  \scaleto{%
    \scalerel*[\widthof{\ensuremath{#1}}]{\kern-.6pt\bigwedge\kern-.6pt}%
    {\rule[-\textheight/2]{1ex}{\textheight}}
  }{\textheight}%
}{0.5ex}}%
\stackon[1pt]{#1}{\scalebox{-1}{\tmpbox}}%
}

\newcommand{\Ab}{\hat{A}}

\newcommand{\alphab}{\hat{\alpha}}
\newcommand{\alphac}{\check{\alpha}}

\newcommand{\Gammab}{\hat{\Gamma}}

\newcommand{\DPsi}{\mathcal{D}\Psi}
\newcommand{\Dpsi}{\mathcal{D}\psi}

\newcommand{\Lc}{\mathcal{L}}
\newcommand{\Ac}{\mathcal{A}}
\newcommand{\rv}{\vec{r}}

\newcommand{\nv}{\vec{n}}

\newcommand{\uv}{\vec{u}}
\newcommand{\kv}{\vec{k}}

\newcommand{\rvp}{\vec{r}\;'}

\newcommand{\Jv}{\vec{J}}

\newcommand{\PP}{\mathbb{P}}

\newcommand{\Psib}{\hat\Psi}
\newcommand{\delb}{\hat\delta}
\newcommand{\sigmab}{\hat\sigma}
\newcommand{\epsb}{\hat\epsilon}

\newcommand{\sigmabbar}{{\hat{\overline \sigma}}}
\newcommand{\sigmabar}{\overline \sigma}
\newcommand{\pbar}{\overline p}
\newcommand{\psibar}{{\overline \psi}}

\newcommand{\fsl}[1]{\ensuremath{\mathrlap{\!\not{\phantom{#1}}}#1}}

\begin{document}
\title{Edwards field theory for glasses and granular matter}
\author{E. DeGiuli}
\affiliation{Institut de Physique Th\'eorique Philippe Meyer, \'Ecole Normale Sup\'erieure, \\ PSL University, Sorbonne Universit\'es, CNRS, 75005 Paris, France}

\begin{abstract}
A minimal description of the inherent states of amorphous solids is presented. Using field theory, applicable when a system is probed at long length scales, it is shown that athermal amorphous solids have long-range correlations in their stresses, as recently observed in supercooled liquids, colloids, and granular matter. Explicit predictions for the correlators are presented, in both 2D and 3D, in excellent agreement with simulation data on supercooled liquids. It is shown that when applied to solids with strictly repulsive interactions, the simplest, na\"ive theory leads to a paradox. This paradox is resolved, and it is shown that a nontrivial, non-Gaussian theory is necessary for such materials. Modifications to the correlators are shown, at the saddle-point level. In all cases, `equations of state' relating fluctuations to imposed stresses are derived, as well as field equations that fix the spatial structure of stresses in arbitrary geometries. A new holographic quantity in 3D amorphous systems is identified. 
\end{abstract}
\maketitle

Amorphous solids have degrees-of-freedom (DOF) in their inherent states that have no counterpart for perfect crystals. Both for glasses at low temperature, and for out-of-equilibrium athermal solids like granular matter, foams, and emulsions, precise characterization of these inherent states remains an unsettled problem. While microscopic details of inherent states will vary from material to material, certain macroscopic properties may be universal. Recent evidence for universality comes from two fronts.

First, there have been many observations of long-range stress correlations in amorphous solids. In simulations both of model granular materials \cite{Henkes09a,Wu17}, model glasses \cite{Wu17} and deeply supercooled liquids  \cite{Lemaitre14,Wu15,Lemaitre15}, the spatial shear-stress correlator has quadrapolar anisotropy and a power-law decay $\propto 1/r^d$ in $d$ dimensions. Strain correlations measured experimentally show similar behavior, both for colloidal glasses \cite{Chikkadi11,Jensen14,Illing16} and granular materials \cite{Le-Bouil14}. In anisotropic photoelastic-disk packings, a model granular material, direct measurements also evidence long-range stress correlations \cite{Bi13,Sarkar13}. 

Second, a large research effort is devoted to understanding the plasticity and eventual yielding of amorphous solids. It is accepted that plasticity is initiated by local instabilities, deemed shear-transformations, which can then, through long-range elastic coupling, trigger further instabilities, leading to avalanches of plastic activity. This paradigm has been explored in detail in computer simulations of glass-formers \cite{Karmakar10a,Lemaitre09,Maloney04,Maloney06a}, and is supported by experimental measurements on colloidal glasses \cite{Chikkadi11,Jensen14}, granular media \cite{Amon2012,Le-Bouil14}, and emulsions \cite{Hebraud97}. One strand of research is dedicated to prediction of localized instabilities from particle positions \cite{Gartner16,Zylberg17}. A second strand of work aims to predict the mesoscopic avalanche properties and macroscopic plastic flow curves \cite{Lin14,Lin14a,Nicolas17}. 

The relation between long-range stress correlations and localized shear-transformations is not entirely straightforward. On one hand, Lema\^itre showed that elastic relaxation of so-called Eshelby transformations is sufficient to explain long-range stress correlations \cite{Lemaitre14,Lemaitre15}. On the other hand, for granular matter the elastic range is extremely small, such that essentially all observed deformation is plastic \cite{Roux10,Schreck11}, casting doubt on this explanation when applied to these materials. A promising alternative is to argue that mechanical equilibrium is sufficient to explain both the long-range stress correlations and localized activity \cite{Lemaitre17}. In this work we present a theory aimed in this direction.

We will present a field theory of inherent structures, which we refer to as an `Edwards field theory.' This references early contributions of Edwards \cite{Edwards05,Edwards89}, who sought to establish a basis for a theory of granular matter. Edwards' proposal was to consider the uniform ensemble over all metastable states, {\blue now known as an Edwards ensemble, } which in principle can be treated by the methods of statistical physics \cite{Bi15}. This simple statement belies two great difficulties, which have preoccupied the field in the past decades: first, establishing the relevant macroscopic control parameters that characterize metastable states; second, actually performing the highly constrained sum over metastable states. 

Initially, many works considered the volume ensemble of $N$ hard grains in a volume $V$ \cite{Edwards89,Makse02,Metzger04,Asenjo14}; later, the stress tensor $\sigmab$ was added as a relevant variable \cite{Edwards05,Henkes05,Henkes07,Henkes09,Henkes09a,Blumenfeld09,Chakraborty10,Blumenfeld12,Wang12,Puckett13,Bi13,Sarkar13,Bililign18}. Despite some empirical successes \cite{Barrat00,Biroli01a,Makse02,Henkes07,Puckett13} the approach has remained controversial, because the assumption of a flat measure has never been justified. Moreover, even assuming the flat measure as a theoretical starting point, the sum over metastable states is extremely difficult to perform. Significant effort has been expended in finding alternative representations of relevant DOF in order to work on the manifold of metastable states \cite{Satake04,Satake86,Satake92,Satake93,Satake97,Ball02,Henkes09,DeGiuli11,DeGiuli13,DeGiuli14a}, {\blue and exact computations have been performed in the limit of infinite dimensions \cite{Charbonneau17}, but so far no exact computation of a genuine Edwards ensemble has been performed in physical dimensions}. 

In this work we attempt to overcome these difficulties. First, we abandon the proposal of a flat ensemble, which was never well justified. In order to constrain the theory, we instead restrict consideration to probing at small wavenumber $k D \ll 1$, where $D$ is a typical particle diameter; then standard methods of statistical field theory can be used to determine which interactions are relevant. Second, following pioneering work by Henkes and Chakraborty on 2D granular matter \cite{Henkes05,Henkes07,Henkes09,Henkes09a}, we work in the continuum by constructing a field theory. In doing so, it is possible to work directly on the manifold of metastable states, and see the nontrivial consequences arising from this restriction. 

We will not restrict attention to granular matter, but treat in a common framework all athermal amorphous solids with finite-range interactions in 2 and 3 dimensions. Initially, we consider generic glasses with both attractive and repulsive interactions, and then derive the additional features in solids with strictly repulsive interactions. Our results shed light on several of the above issues. Namely, we will predict
\begin{itemize}
\item Athermal amorphous solids have long-range stress correlations. Existence of correlations follows from mechanical equilibrium alone. An explicit formula for the stress correlator will be derived in field theory, both for 2 and 3 dimensions. These results are in excellent agreement with simulations of supercooled liquids.
\item Equations of state relate external control parameters to applied stresses, explicit forms of which are presented. 
\item Field equations govern the spatial distribution of stress in arbitrary geometries. These will be derived, along with appropriate boundary conditions.
\item A new controllable quantity in 3D systems will be presented, the Beltrami volume. 
\item For solids with strictly repulsive interactions, the equation of state and field equations are modified. Connected stress correlators are nontrivial at all orders. The pressure field then has long-range correlations, which may explain recent results \cite{Wu17}.
\end{itemize}

A brief account of these results has been presented in \cite{atmp_DeGiuli18_short}. 

Our tensor notation is such that all contractions are explicitly indicated. We alternatively use index-free notation, when appropriate, and indices when necessary, with the Einstein convention. The identity tensor is denoted $\delb$. We let tensor divergences act on the first index, e.g. $\nabla \cdot \hat{A} = \p_i A_{ij}$, while, by convention, the tensor curl operates on the final index, e.g. $\nabla \times \hat{A} = \epsilon_{ijk} \p_j A_{lk}$. In 2D, we make frequent use of the antisymmetric tensor, $\epsilon_{12}=-\epsilon_{21}=1, \epsilon_{11}=\epsilon_{22}=0$. {\blue We recall that $\epsb^{-1} = -\epsb = \epsb^t$. In 2D, it is useful to introduce an inverted-hat notation, } for tensors,
\eq{
\check{A} \equiv \epsb^t \cdot \Ab \cdot \epsb = \delb \;\tr \Ab - \Ab^t,
}
where the last equality follows from $\epsilon_{ij} \epsilon_{kl} = \delta_{ik} \delta_{jl} - \delta_{il} \delta_{jk}$. {\blue We will frequently use the simple relation
\eq{
\check{A} : \hat{B} = \hat{A} : \check{B}
}
}

\section{Stress correlations in mechanical equilibrium}

Inherent states are defined by conditions of mechanical equilibrium. Unlike crystals, for which these constraints are trivially satified by symmetry, in amorphous materials mechanical equilibrium imposes nontrivial constraints on the microscopic DOF. In this work, we consider the constraints only on stress; microscopically, forces and particle positions are of course coupled, but we can marginalize over the geometric DOF to work in the reduced `stress ensemble' \cite{Henkes09a,Chakraborty10}. We suppose that this marginalization is benign, in the sense that no long-range interactions are induced by marginalization. Future work will examine the circumstances under which this assumption is correct. 

Before constructing a complete theory of the stress ensemble, it is instructive to see how the conditions of mechanical equilibrium strongly constrain the tensorial form of correlation functions, without any additional hypotheses. This point has recently been vividly demonstrated by Lema\^itre in 2 dimensional systems \cite{Lemaitre17}; here we will show how a gauge formulation of the problem immediately gives a compact and complete answer, and then generalize this result to 3 dimensional systems. 

In the absence of body forces, the stress tensor of a system in mechanical equilibrium must be symmetric, $\sigmab=\sigmab^t$, from torque balance, and solenoidal, $0=\nabla \cdot \sigmab$, from force balance. In two dimensions these equations are identically solved by Airy's representation
\eq{ \label{airy}
\sigmab = \nabla \times \nabla \times \psi, \qquad \sigma_{ik} = \epsilon_{ij} \epsilon_{kl} \p_j \p_l \psi,
}
where $\psi$ is known as the Airy stress function \cite{Muskhelishvili63} and $\epsilon_{12}=-\epsilon_{21}=1, \epsilon_{11}=\epsilon_{22}=0$. It is easily verified that for any function $\psi(\rv)$, both $\sigmab=\sigmab^t$ and $0=\nabla \cdot \sigmab$ are identically satisfied. Moreover, this representation also exists at the particle scale \cite{Ball02,Henkes09,DeGiuli11,DeGiuli14a}. The price of the gauge representation is that stresses are invariant under the gauge transformation $\psi \to \psi + \vec{a} \cdot \rv + b$, for any constants $\vec{a}$ and $b$. Stresses depend only on the curvature of $\psi$. 

The fundamental correlation function is therefore
\eq{
C_\psi(\rv,\rvp) = \langle \psi(\rv) \psi(\rvp) \rangle_c,
}
which is invariant under the gauge transformation $\psi \to \psi + \vec{a} \cdot \rv + b$. The 
 stress-stress correlation function is then
\eq{
\langle \sigma_{ij}(\rv) \sigma_{kl}(\rvp) \rangle_c & = \epsilon_{im} \epsilon_{jn} \epsilon_{kp} \epsilon_{lq} \p_m \p_n \p'_p \p'_q C_\psi(\rv,\rvp)
}
Assuming homogeneity, this can be written
\eq{ \label{corr3}
\langle \sigma_{ij}(\rv) \sigma_{kl}(0) \rangle_c & = \epsilon_{im} \epsilon_{jn} \epsilon_{kp} \epsilon_{lq} \p_m \p_n \p_p \p_q C_\psi(\rv,0).
}
Since the pressure-pressure correlator is $\langle p(\rv) p(0) \rangle_c = \ffrac{1}{4} \nabla^4 C_\psi(\rv,0)$, in periodic systems the full correlation function can be written
\eq{ \label{corr4}
\langle \sigma_{ij}(\rv) \sigma_{kl}(0) \rangle_c = 4 P_{ij}^T P_{kl}^T \langle p(\rv) p(0) \rangle_c,
}
where $P_{ij}^T$ is the transverse projector \cite{DiDonna05}, which in Fourier space is simply
\eq{
P_{ij}^T = \epsilon_{im} \epsilon_{jn} \frac{k_m k_n}{k^2} = \delta_{ij} - \frac{k_i k_j}{k^2} 
}
Eq.\ref{corr4} holds even in anisotropic systems; the tensorial structure is entirely fixed by mechanical equilibrium.




From \eqref{corr3} one can easily determine all components of the stress correlator if $C_\psi(\rv,0)$ is known. It will be the object of later sections to derive the latter. But let us first extend the above results to three dimensions.

On simply connected domains\footnote{In multiply connected domains, topological excitations require additional harmonic terms in $\sigmab$. See \cite{Wang13a}}, the representation analogous to \eqref{airy} in 3D is 
\eq{ \label{beltrami}
\sigmab = \nabla \times \nabla \times \Psib, \qquad \sigma_{il} = \epsilon_{ijk} \epsilon_{lmn} \p_j \p_m \Psi_{kn},
}
where $\Psib$, a symmetric second-order tensor, is the Beltrami stress tensor \cite{Gurtin63}. Note that, by convention, the tensor curl is defined by acting on the right-most index, i.e. $(\nabla \times \Psib)_{ij} = \epsilon_{ikl} \p_k \Psi_{jl}$. A discrete representation of $\Psib$ also exists \cite{DeGiuli11}. What is the gauge freedom of $\Psib$? For any vector field $\vec{p}(\rv)$, the stress tensor is invariant under the transformation $\Psib \to \Psib + \nabla \vec{p} + (\nabla \vec{p})^t$,
which thus constitutes a nontrivial gauge group \cite{Wang13a}. Accordingly, $\Psib$ has redundant degrees of freedom, and can be further reduced. In the literature, one finds the Morera gauge, where $\Psi_{ij}=0$ if $i=j$, and the Maxwell gauge, where $\Psi_{ij}=0$ if $i \neq j$ \cite{Gurtin63}. Completeness of both representations has been proven for sufficiently smooth stress fields \cite{Gurtin73,Rostamian79}. We will use the Maxwell gauge $\Psi_{ij} = \delta_{ij} \psi_j$ (no sum on $j$), with a residual gauge group $\psi_j \to \psi_j + p_j(\rv)$, with $\p_k p_j=0$, $j \neq k$ (thus $p_j$ cannot be rotationally symmetric). Then the fundamental correlation function is
\eq{
C_{ij}(\rv,\rvp) = \left\langle \psi_i(\rv) \psi_j(\rvp) \right\rangle_c,
}
which has at most 6 independent components, and is gauge invariant. If isotropy and homogeneity are assumed, then this has two independent components, $A(\rv)=C_{ii}(\rv,0)$ (no sum on $i$) and $B(\rv)=C_{ij}(\rv,0)$ ($i\neq j$). {\blue As shown in Appendix 1, all stress correlators} involving longitudinal components vanish, so that again only the transverse-transverse stress correlator survives, which now, however, is tensorial. 

Although simple to derive in the gauge formulation, the above results completely prescribe the tensorial structure of the stress correlator, a major aim of previous works \cite{Lemaitre14,Lemaitre15,Maier17,Lemaitre17} . We also see that material isotropy is not important in determining this structure, although it would simplify the implied derivatives. To obtain predictions for the correlation functions, we now proceed to statistical mechanics.

\section{Gauge field theory of inherent states}
\subsection{Ensembles for athermal systems}

{\blue Here we review and extend general arguments for construction of a statistical ensemble in systems where the Boltzmann-Gibbs distribution does not necessarily apply \cite{Edwards05,Bertin06,Henkes07,Henkes09a,Tighe11a,Wu15a}. } We are interested both in glasses and out-of-equilibrium athermal systems. For glasses, the probability distribution over inherent states will contain a Gibbs contribution from the energy at the glass transition temperature, but also an entropic contribution, the `complexity.' \cite{Berthier11b,Charbonneau17}. The latter is highly nontrivial and can depend on all the parameters of the system. 
 For athermal systems, we do not even have a Gibbs contribution from which to begin a theory. 
 
To construct a stress ensemble valid out of equilibrium, we will therefore take an operational point of view: {\blue typically, physical three-dimensional systems can only be probed through forcing at the boundary. } Unlike thermally equilibrated systems, an athermal ensemble needs to be explicitly explored through systematic forcing. Such an ensemble can be explored dynamically, as in quasistatic shear flow in a Couette cell, but we need not restrict ourselves to this setting; indeed, most numerical simulations and experiments simply use repeated application of a preparation protocol. {\blue One can of course imagine ensembles created by means other than boundary forcing, especially in numerical simulation, such as by repeated local probes. The difficulty with such ensembles is that they allow Maxwell demons.  
The procedure we follow is analogous to construction of classical thermal statistical mechanics, as discussed more below. } 


 In order for a variable to be controllable under an athermal ensemble generated by boundary forcing, it must be {\it holographic,} that is, determined by boundary quantities only. The stress tensor in a mechanically equilibrated system is indeed such a quantity, as can be seen by taking a tensorial moment of the condition $\nabla \cdot \sigmab = 0$:
\eq{
0 = \int_\Omega dV  \;(\nabla \cdot \sigmab) \vec{r} = \int_\Omega dV \;\nabla \cdot \big( \sigmab \vec{r} \big) - \int_\Omega dV \; \sigmab,
}
where our tensor notation is such that all tensor contractions are explicitly indicated by dots. In the final equation, the divergence theorem implies that $\int dV \nabla \cdot (\sigmab \vec{r}\;)$ is a boundary term, while the second term gives the volume integral of the stress tensor, hence the latter is holographic. This result also holds at the particle scale \cite{Kruyt96}. 

In addition to being holographic, controllable quantities should be additive, so that the thermodynamic limit can exist. Thus the true controllable quantity is $\int_\Omega dV \sigmab$, also known as the force-moment tensor. {\blue This should be compared to the behavior of energy in classical statistical mechanics. Since energy is conserved in time, its integral over time can written as boundary terms involving initial and final times. Energy is controllable in classical statistical mechanics in the same sense as force-moment above\footnote{Note also that classical statistical mechanics of conservative systems, with nontrivial boundary conditions only in time, and Edwards statistical mechanics, as defined here without time but with nontrivial boundary conditions in space, are end members of a larger class of theories in which boundary conditions can be nontrivial in space and time. For example, Galley has shown that Lagrangian theories with initial conditions in time, rather than boundary conditions, describe the dynamics of dissipative systems \cite{Galley13}.}.}

It is a surprising fact that in addition to $\int_\Omega dV \sigmab$, there is another holographic, additive quantity depending on the stress \cite{Tighe08,Tighe10a,Tighe10b}. To see this, we initially consider two dimensions, and use the Airy representation \eqref{airy}.  
 One sees that the determinant of the stress tensor is then
\eq{
\det \sigma & = \half \epsilon_{ij} \epsilon_{kl} \sigma_{ik} \sigma_{jl} = \half \epsilon_{ij} \epsilon_{kl} \epsilon_{im} \epsilon_{kn} (\p_m \p_n \psi) \sigma_{jl}  \notag \\
& = \half \p_j \big( (\p_l \psi) \sigma_{jl} \big)
}
where we used $\epsb^T \cdot \epsb = \delb$ and $\nabla \cdot \sigmab = 0$. 
Thus $\Ac = \int_\Omega dV \det \sigma$ can be written as a boundary quantity. Note that since $\psi$ has a gauge freedom $\psi \to \psi + \vec{a} \cdot \rv + b$, the flux 
$J_j = (\p_l \psi) \sigma_{jl}$
 has a gauge-dependent solenoidal part. This does not affect the obviously gauge-independent $\det \sigma = \half \p_j J_j$. In previous work, the discrete quantity corresponding to $\Ac$ has been called the Maxwell-Cremona area \cite{Tighe08,Tighe10a,Tighe10b,DeGiuli11,Wu15a}. 

Let us now show that a similar quantity also exists in 3 dimensions, although to our knowledge it has never been reported before. Using the gauge representation \eqref{beltrami}, simple algebra (Appendix 1) shows that the determinant of $\sigmab$ is now
\eq{
\det \sigma & = \frac{1}{3} \epsilon_{lmn} \p_p \left[ (\nabla \times \Psib)_{lq} \sigma_{pm} \sigma_{qn} \right],
}
which is a total divergence. The quantity $\Ac = \int_\Omega \det \sigma$ could be called the Beltrami volume. 
 As in 2D, the flux 
$J_p = \epsilon_{lmn} (\nabla \times \Psib)_{lq} \sigma_{pm} \sigma_{qn}$ 
 has a gauge-dependent solenoidal part. 

Having identified controllable quantities $\int_\Omega \sigma$ and $\Ac$, we can construct a canonical ensemble in which the control parameters are temperature-like variables conjugate to $\sigma$ and $\Ac$. This leads to an action
\eq{
S_0 = \int_{\Omega} dV \; \left[ \alphab : \sigmab + \gamma \det \sigmab \right],
}
where $\alphab^{-1}$ has been called the angoricity \cite{Blumenfeld09}, and $\gamma$ has been called the keramicity \cite{Bililign18}. The justification for the canonical ensemble is based upon an assumed factorization of the probability distribution for macroscopic variables into that of subsystems, and is discussed in detail in Refs. \cite{Bertin06,Henkes07,Henkes09a}. In such a generalized Gibbs ensemble, the temperature-like variables $\alphab$ and $\gamma$ are argued to be spatially constant \cite{Bertin06}.

To complete the specification of the probability distribution of the stress field, we need to address (i) the hard constraints necessary to impose mechanical equilibrium, and (ii) the {\it a priori} probability with which each metastable state is sampled. Since $\sigmab=\sigmab^t$ and $0 = \nabla \cdot \sigmab$ are identically solved by the Airy (2D) and Beltrami (3D) stress functions, we can efficiently work on the manifold of metastable states by writing $\sigmab$ as a functional of $\psi$ (2D) and $\psi_i$ (3D). This leads to
\eq{ \label{p1}
\PP[\sigmab[\psi]] = \frac{1}{Z} \omega[\sigmab[\psi]] e^{-S_0[\psi]},
}
where $\omega$ is the sampling probability of the state defined by $\sigmab[\psi]$, and we use $\psi$ to refer either the scalar Airy stress function (2D) or its vectorial analog in the Maxwell gauge (3D). It is implicit that in $\psi-$space there is a UV cutoff $\Lambda \propto 1/D$, where $D$ is the typical particle diameter.

In a strict canonical ensemble, the sampling probability $\omega$ would be unity, as was taken in previous work on the stress ensemble \cite{Henkes07,Henkes09}, although it was recognized that the statistical mechanical formalism does not require this \cite{Chakraborty10}. In fact, there is no general justification for the flat measure out of equilibrium, even if it was observed to hold to a good approximation in several model systems \cite{Barrat00,Biroli01a}.  In general, we expect the flat measure to be unrealistic for a simple reason: since $S_0$ can be written in terms of boundary quantities only, if $\omega\equiv 1$ then \eqref{p1} would be invariant under arbitrary diffeomorphisms in the bulk, limited only by the UV cutoff $\Lambda$. This would allow arbitrarily wild fluctuations of the field down to the scale $\Lambda$, which is not physical: a solid stores elastic energy, and whenever elasticity is present, stress fluctuations will be penalized. 

The sampling probability $\omega$ must thus be nontrivial. Initially this looks hopeless, because for arbitrary $\omega$ nothing can be computed, but we are rescued by the continuum limit. The general theory of the renormalization group indicates that when a system is probed at long length scales, most of its microscopic details are irrelevant \cite{Zee10,Zinn-Justin96}. For any theory with a Lagrangian, power counting can be applied to see which terms are necessary to retain in a general expansion
\eq{
\omega[\sigma[\psi]] = e^{-\int dV [ A_1[\sigmab] + A_2[\sigmab,\sigmab] + \ldots ]},
}
where each $A_i$ is a differential operator linear in each argument. Tacitly we are assuming that $\log \omega$ contains only simple powers of $\sigmab$ and their derivatives, the usual Landau expansion. We will return to this point below, when we discuss granular matter.

\subsection{Solids with both attractive and repulsive interactions}

To constrain $\omega$, we need to consider the symmetry properties of the stress tensor. Here we consider systems with both repulsive and attractive interactions; systems with only repulsive interactions are considered in II C. With both attraction and repulsion a term linear in stress, which is not invariant under $\sigmab \to -\sigmab$, will not ensure a well-behaved distribution; for this a term quadratic in stress is necessary. In the continuum limit, the lowest order term necessary to tame fluctuations is then $\eta \;\sigmab : \sigmab$. We assume that $\eta$ defines the correct units in which to construct the field theory, {\blue meaning that the term $\int dV \eta \sigmab : \sigmab$ survives in the continuum limit for any value of $\eta$. Since the action is dimensionless, if we assign a dimension $+1$ to lengths and a dimension 0 to $\eta$, then $\sigmab$ must have canonical dimension \footnote{Equivalently, we could define a rescaled $\sigmab' = \sqrt{\eta} \sigmab$ and find the dimension of $\sigmab'$ necessary to make $\int dV \sigmab' : \sigmab'$ dimensionless.} $-d/2$. } A term of the form $\p^n \sigma^q$ then has a coupling constant with operator dimension $\delta_{n,q}=d-n-qd/2$. Relevant operators are those with $\delta_{n,q} \geq 0$, since these will, at least perturbatively, stay finite in the continuum limit $kD \to 0$ \cite{Zee10}. In $d=2,3$ this includes only $q=1,n \leq1$, and $q=2,n=0$. Assuming reflection symmetry (no gravity), so that a term $g_{ijk} \p_i \sigma_{jk}$ is excluded, the only new isotropic terms added are $\eta \;\tr^2\sigmab$ and $g \;\tr\; \sigmab^2$. The leading anisotropic term is $\alpha_{ij} \sigma_{ij}$, which we retain. Higher-order anisotropic terms are possible, such as $g_a F_{ij} \sigma_{ij} \sigma_{kk}$, and under strongly anisotropic forcing, such terms would be necessary; {\blue this is discussed more below. }

We are thus led to consider
\eq{ \label{p2}
\PP[\sigmab[\psi]] = \frac{1}{Z} e^{-S[\psi]}, \;\; S = \int_{\Omega} dV \; \Lc[\psi], 
}
with
\eq{ \label{l1}
\Lc[\psi] & = \alphab : \sigmab + \gamma \det \sigmab + \half \eta \; \tr^2 \sigmab + \half g \; \tr\;\sigmab\cdot \sigmab.
}
Both $\eta$ and $g$ should be positive to suppress fluctuations. As usual, it is sufficient to compute $Z = \int \Dpsi \;e^{-S}$ to extract the behavior of controllable quantities. For example, writing $\overline{x} \equiv \ffrac{1}{|\Omega|} \int_\Omega dV \; x(\rv)$ for a spatial average one easily sees that
\eq{
& \langle \overline{\sigmab} \rangle = - \frac{1}{V} \frac{\p \log Z}{\p \alphab}, \qquad \langle \overline{\sigmab \sigmab} \rangle_c = \frac{1}{V^2} \frac{\p^2 \log Z}{\p \alphab \p \alphab}, \\
& \left\langle \overline{\tr^2 \sigmab} \right\rangle = -\frac{2}{V} \frac{\p \log Z}{\p \eta},
}
where $V$ is the system volume.

At this stage, it is clear that we could have arrived at \eqref{l1} with power counting alone, without any consideration of controllable quantities. However, this would miss an important point: the parameters $\alphab$ and $\gamma$ are conjugate to holographic quantities, and hence in principle under experimental control. The `elastic' parameters $\eta$ and $g$ instead reflect the properties of the particles and should not depend on details of the experimental protocol. 


We will see further the importance of this distinction below. Note that in 3D this distinction between parameters is precise, but in 2D, there is an ambiguity, because $\det \sigmab = \half (\tr^2 \sigma - \tr \sigmab^2)$, so that $\gamma, \eta,$ and $g$ are not independent; we will absorb $g$ into $\tilde \eta = \eta +g$, $\tilde \gamma = \gamma - g$.


{\bf Infinite shear symmetry: } Remarkably, \eqref{l1} has an infinite dimensional symmetry. In 2D and 3D, respectively, consider the transformation 
\eq{ 
\psi(\rv) & \to \psi(\rv) + h(\rv) & \qquad \mbox{2D} \label{sym2d} \\
\Psib(\rv) & \to \Psib(\rv) + \delb \; h(\rv) & \qquad \mbox{3D} \label{sym3d}
}
where $h(\rv)$ is harmonic, $\nabla^2 h(\rv) = 0$, but otherwise arbitrary. It is straightforward to compute that \eqref{sym2d},\eqref{sym3d} leaves the pressure $p$ invariant, and $\Lc$ itself invariant up to a boundary flux, so that all such stress changes are symmetries of the action. These are gauge symmetries of \eqref{p2}, but with a clear physical meaning: {\blue since they preserve the pressure, they are increments of shear stress. }
 This symmetry is expected to play an important role in the dynamics near inherent states, to be considered in future work.

{\bf 2D: } We now consider dimensions 2 and 3 separately. In 2D. the partition function to be computed is
\eq{
Z = \int \Dpsi \;e^{-S}, \qquad\;\; S = \int_{\Omega} dV \; \Lc[\psi], 
}
with
\eq{
\Lc[\psi] & = \alphab : \sigmab + \gamma \det \sigmab + \half \eta \; \tr^2 \sigmab \\
& = \alphab : (\epsb^t \cdot \nabla \nabla \psi \cdot \epsb) + \tilde\gamma \det \nabla \nabla \psi + \half \tilde\eta \; (\nabla^2 \psi)^2 \notag,
}
where $\tilde \eta = \eta +g$, $\tilde \gamma = \gamma - g$. {\blue In the physical system, $\alphab, \gamma, \eta,$ and $g$ are constant in space. However, correlation functions can be generated by allowing $\alphab$ to be space-dependent \cite{Zee10}. We let } $\alphab = \hat{\overline{\alpha}} + \alphab_g(\rv)$ where the former controls the mean stress and the latter generates correlation functions. One easily sees that
\eq{ \label{corr0}
\langle \sigmab(\rv) \sigmab(\rvp) \rangle_c 
= \frac{\delta^2 \log Z}{\delta \alphab_g(\rv) \delta \alphab_g(\rvp)},
}
At the end of the computation, one can then set $\alphab_g=0$ to recover the physical ensemble.

It is useful to note that $\psi$ is not translationally invariant. For example, the solution to $\sigmabar = \nabla \times \nabla \times \psibar$ is 
\eq{
\psibar = \half \rv \times \sigmabar \times \rv = \half r_i \epsilon_{ij} \sigmabar_{jk} \epsilon_{kl} r_l,
}
which grows as $r^2$. For this reason, in computing $Z$ we cannot disregard boundary terms with inpunity. The equation-of-state will in fact be determined by boundary fluxes. 

Since $\Lc$ is quadratic in the field $\psi$, $Z$ can be computed exactly in the continuum limit. It is convenient to use the `background field method,' where we let $\psi = \psi_c + \psi'$ and choose $\psi_c$ to eliminate cross-coupling between $\psi_c$ and $\psi'$. This procedure avoids complications from functional integrations, and generalizes well to nonlinear situations. The result in the thermodynamic limit  (Appendix 2) is $\log Z = - S_c - S_p$ with
\eq{
S_c = \half \int dV \; \alphab : \sigmab_c, \qquad S_p = \frac{V\Lambda^2}{8\pi}  \log \frac{\tilde\eta \Lambda^4}{e^2},
}
where the `classical' part $\psi_c$ must solve
\eq{ \label{eom1}
\nabla^4 \psi_c = -\tilde\eta^{-1} \nabla \nabla : \alphac
}
with boundary conditions
\eq{ \label{bc1}
0 & = \nv \cdot \left[ \alphac + \tilde \gamma \sigmab_c +  \tilde\eta \; \delb (\nabla^2 \psi_c) \right], \notag \\
0 & = \nv \cdot \left[ \nabla \cdot \alphac + \tilde\eta \nabla (\nabla^2 \psi_c) \right]
}
where $\nv$ is a boundary normal, and we use the notation $\check{A} \equiv \epsb^t \cdot \Ab \cdot \epsb = \delb \;\tr \Ab - \Ab^T$.
 We write $\psi_c = \psibar + \psi_{g}$ with $\psibar$ as above. We see that in order to cancel the term $\check{\overline{\alpha}}$ in the boundary conditions, $\sigmabbar$ must satisfy $0 = \check{\overline{\alpha}} +  \tilde\gamma \sigmabbar +  \tilde\eta \delb \;\tr\; \sigmabbar$, which leads to  
\eq{ \label{eos1}
\sigmabbar = \frac{1}{\gamma-g} \hat{\overline{\alpha}} - \frac{(\eta+\gamma)}{(\gamma-g)(\gamma + g + 2 \eta)} \delb \;\tr \;\hat{\overline{\alpha}}.
}
This is the equation of state relating the temperature-like quantities $\alphab$ and $\gamma$ to the mean stress $\langle \sigmab \rangle = \sigmabbar$. If boundary terms had been neglected from the outset, we would not have obtained this equation. The inhomogeneous part $\psi_g$ must satisfy $\nabla^4 \psi_g = -\tilde\eta^{-1} \nabla \nabla : \alphac_g$ and the boundary conditions \eqref{bc1} with $\alpha \to \alpha_g$. In an infinite domain, the solution to a source $\alphab_g = \alphab_0 \delta(\rv)$ is
\eq{ \label{corr5}
4 \pi \tilde\eta \; \psi_g = -\alpha \log r^2 + a \cos 2\theta + b \sin 2\theta, 
}
where 
\eq{
\alphab_0 = \begin{pmatrix} \alpha + a & b \\ b & \alpha-a \end{pmatrix}
}
Boundary conditions can be applied by adding to $\psi_g$ an appropriate biharmonic function $\psi_b$, $\nabla^4 \psi_b=0$. 
 As noted above, $\Lc$ has a large symmetry: any harmonic function $h(\rv)$ can be added to $\psi$ while only changing the action by a boundary term. This symmetry is reflected here in the ability to add a biharmonic function $\psi_b$ to $\psi_g$. 

The solution to a general collection of sources $\alphab_g(\rv') = \sum_i \alphab_i \delta(\rv-\rv_i)$ is then obtained by superposition:
\eq{ \label{corr6}
\psi_g(\rv) & = \frac{-1}{4\pi \tilde\eta} \sum_i \; \left[ \alpha_i \log |\rv-\rv_i|^2 \right. \notag \\
& \qquad \left. - a_i \cos 2\theta_{rr_i} - b_i \sin 2\theta_{rr_i} \right],
}
where $\theta_{rr_i}$ is the polar angle of $\rv-\rv_i$. 

Since the `phonon' part $S_p$ does not depend on $\alphab$, the correlation function is
\eq{ \label{corr7}
\langle \sigmab(\rv) \sigmab(\rvp) \rangle_c = -\half \delta \sigmab_g(\rv) / \delta \alphab_g(\rvp) -\half \delta \sigmab_g(\rvp) / \delta \alphab_g(\rv), 
}
evaluated at $\alphab_g=0$. For example, the pressure-pressure correlator is
\eq{
\langle p(\rv) p(0) \rangle_c = \frac{1}{4\tilde\eta} \; \delta(\rv)
}
This is short-range, but all second derivatives will have a $1/r^2$ decay with appropriate anisotropic dependencies, following Eq.\eqref{corr4}. Note that the prediction of a perfect $\delta(\rv)$ correlator is an artifact of the truncation of $\Lc$ to Gaussian order; if higher order terms were included in $\Lc$, such as $\tr^4 \sigma$, then the pressure-pressure correlator would have an exponential decay over the particle size  length scale $\sim D$, as observed in \cite{Henkes09a}.

The correlator for $\psi$ corresponding to \eqref{corr5} is
\eq{
C_\psi(\rv,0) = \frac{r^2 \log r}{4 \pi \tilde\eta}
}
up to irrelevant terms that do not affect the stress correlator. From this we can compute, for example,
\eq{
\langle \sigma_{xx}(\rv) \sigma_{xx}(0) \rangle_c & = \frac{3}{8\tilde\eta} \delta(\rv) + \frac{2 \cos 2\theta_r + \cos 4\theta_r}{4 \pi \tilde\eta r^2}, \\
\langle \sigma_{yy}(\rv) \sigma_{yy}(0) \rangle_c & = \frac{3}{8\tilde\eta} \delta(\rv) + \frac{- \cos 4\theta_r}{4 \pi \tilde\eta r^2}, \\
\langle \sigma_{xy}(\rv) \sigma_{xy}(0) \rangle_c & = \frac{1}{8\tilde\eta} \delta(\rv) + \frac{-2 \cos 2\theta_r + \cos 4\theta_r}{4 \pi \tilde\eta r^2},
}
In addition to the $1/r^2$ decay discussed above, several characteristic properties of this solution were observed in previous work \cite{Lemaitre14}: the ratio of the $\delta(\rv)$ amplitude in the $p$ correlator to that of the $\sigma_{xy}$ correlator is $1/2$, and the ratio of the coefficient of the $\cos 4\theta$ terms to that of the $\cos 2\theta$ terms is also $1/2$. Similar relations can be obtained for the normal stress difference $(\sigma_{xx}-\sigma_{yy})/2$ by symmetry. These results agree with those of Henkes and Chakraborty in the macroscopic limit $qD \to 0$. 

Finally, the field equation $\nabla^4 \psi_c = 0$ applies whenever there are large-scale variations in $\sigma$ arising from boundary conditions. Since this is equivalent to the equation satisfied by $\psi$ in linear elasticity \cite{Sadd09}, solutions will coincide when stresses are fixed on the boundary. To apply boundary conditions on displacements, a proper treatment of geometrical variables is necessary. This is left for future work. 


{\bf 3D}: In three dimensions, we can proceed similarly. The Lagrangian is now 
\eq{
\Lc[\Psi] & = \alphab : \sigmab + \gamma \det \sigmab + \half \eta \; \tr^2 \sigmab + \half g \; \tr\;\sigmab\cdot \sigmab
}
We write $\Psi = \Psi_c + \Psi'$ and find (Appendix 3) that $\Psi_c$ must solve
\eq{ \label{class3d}
0 & = \epsilon_{lmi} \epsilon_{nkj} \p_m \p_k \alpha_{ln} + (\eta+g) \delta_{ij} \nabla^2 \sigma_{c,kk} \notag \\
& \qquad - (\eta+g) \p_i \p_j \sigma_{c,kk} - g \nabla^2 \sigma_{c,ij}
}
subject to boundary conditions
\eq{
0 &= n_k \left[ \alpha_{ij} \epsilon_{ikl}  +  \gamma \; \epsilon_{jmn} \sigma_{c,km} \sigma_{c,ln} +  \eta \; \epsilon_{jkl} \sigma_{c,ii} \right.  \\
& \qquad \left. +  g \; \epsilon_{ikl} \sigma_{c,ij} \right] \notag \\
0 & = n_k \epsilon_{iml} \p_m \left[  \alpha_{ij} \epsilon_{jkn}  + \eta \;\epsilon_{ikn}  \sigma_{c,jj} + g \;\epsilon_{jkn}  \sigma_{c,ij} \right], 
}
where $n_k$ is a boundary normal. We write $\Psi_c = \overline{\Psi} + \Psi_g$, where $\sigmabbar = \nabla \times \nabla \times \overline{\Psi}$ is constant. It is fixed by the equation of state
\eq{
0 = \alpha_{ij} + \eta \; \delta_{ij} \sigmabar_{kk} + g \; \sigmabar_{ij} + \half \gamma \; \epsilon_{ikl} \epsilon_{jmn} \sigmabar_{km} \sigmabar_{ln}
}
As with the 2D case, the partition function separates into its classical and fluctuating contributions. The former is $e^{-S_c}$ with
\eq{ \label{Sc3d}
S_c = \half \int dV \; \left[ \alphab: \sigmab_c - \gamma |\sigmab_c| \right].
}
We have $\sigmab_c = \sigmabbar + \sigmab_g$, with $\sigmabbar$ determined by the equation of state. For a source $\alphab_g=\alphab_0 \delta(\rv)$ at the origin of an infinite domain, $\sigmab_g$ is
\eq{
g \sigmab_g = -\alphab_g - \frac{2\eta}{\tilde g} \delb \; \nabla \cdot \uv + \frac{\eta}{\tilde g} \tr \;\alpha_g + \nabla \uv + (\nabla \uv)^t,
}
with
\eq{
\nabla \cdot \uv & = -\frac{1}{8\pi} \frac{\tilde g}{2\eta+g} \left[ \alphab_0 : \nabla \nabla - \frac{\eta \;\tr\;\alphab_0}{\tilde g} \nabla^2 \right] \frac{1}{r}, \\
\uv & = - \frac{1}{4\pi} \left[ \alphab_0 - \frac{\eta}{\tilde g} \delb \; \tr \;\alphab_0 \right] \cdot \nabla \frac{1}{r}  \notag \\
& \qquad - \frac{\eta+g}{\tilde g} \frac{1}{\nabla^2} \nabla \nabla \cdot \uv.
}
and $\tilde g = 3\eta+g$. The complex tensorial structure resulting from this solution precisely matches what was found in Ref. \cite{Lemaitre15}. For example, the isotropic part is
\eq{
\tr\; \sigmab_g = -\frac{2 \alpha}{2\eta+g} \delta(\rv) - \frac{1}{2\eta+g} \fsl{\alphab} : \nabla \nabla \frac{1}{4\pi r},
}
where $\alphab_0 = \alpha \delb + \fsl{\alphab}$ with $\tr \; \fsl{\alphab} = 0$. The correlator is determined using \eqref{corr0}. By power-counting, the term in \eqref{Sc3d} dependent upon $\gamma$ is expected to be sub-dominant; for simplicity we will neglect it. In this case, the correlator can be determined by Eq.\eqref{corr7}. The pressure-pressure correlator is short-range:
\eq{
\langle p(r) p(0) \rangle_c = \frac{1}{6\eta+3g} \delta(\rv),
}
while the pressure-shear correlator will have anisotropies and long-range decay determined by the Oseen tensor $\nabla \nabla r^{-1}$. 

As in 2D, the classical equation Eq.\eqref{class3d} applies, with $\nabla \alphab=0$, whenever there are large-scale variations in $\sigmab$ arising from boundary conditions. By taking a trace one sees that $\nabla^2 p = 0$ when $\nabla \alphab=0$ so that
\eq{ \label{beltramimichell}
0 = (\eta+g) \p_i \p_j \sigma_{kk} + g \nabla^2 \sigma_{ij}.
}
This is equivalent to the Beltrami-Michell equation of linear elasticity \cite{Sadd09}, with an effective Poisson ratio 
\eq{
\nu = -\eta/(\eta+g)
}

{\bf Holography: } From the above results we can see that the holographic terms play a fundamentally different role from the others. 
 Indeed, $\alphab$ and $\gamma$ appear only in the equations of state, and in boundary conditions for the fluctuations. Since the latter have a negligible effect in large systems, we reach the surprising conclusion that $\alphab$ and $\gamma$ control the system-spanning $\kv=0$ fluctuations, but not the finite wavevector $|\kv|>0$ fluctuations. As a result the stress-stress correlation function should have a discontinuity or kink at $\kv=0$, {\blue which was indeed observed in Ref. \onlinecite{Lemaitre17}. }

To see these distinct fluctuations, let $\overline{x} \equiv \ffrac{1}{|\Omega|} \int_\Omega dV \; x(\rv)$ denote a spatial average, and consider
\eq{
C_e = \left\langle \left(\overline{p - \langle p \rangle} \right)^2 \right\rangle, \qquad C_s = \left\langle \overline{ (p - \overline{p})^2 } \right\rangle,
} 
where $\langle \; \rangle$ denotes an ensemble average. $C_e$ measures the ensemble pressure fluctuations while $C_s$ measures spatial pressure fluctuations. For $d=2$ we find (Appendix 4)
\eq{
C_e = \frac{1}{2 V (2\eta+g+\gamma)}, \qquad C_s = \frac{\Lambda^2}{16 \pi (\eta+g)} - C_e
}
We see that the total fluctuations $C_e+C_s$ are fixed by $\eta$ and $g$ only, while the ensemble fluctuations depend additionally on $\gamma$. This is an expression of the singularity at $\kv=0$.

\subsection{Solids with strictly repulsive interactions: } Previous work on the stress ensemble \cite{Henkes07,Henkes09a,Chakraborty10} has focussed on dry granular material, for which contact forces are strictly repulsive. This constraint, which implies $p>0$ in the continuum, significantly complicates evaluation of $Z$. Let us first see what happens if we try to na\"ively apply the previous results. The local pressure $p(r)$ has a typical magnitude $p(r) \sim \overline{p} \pm \sqrt{C_s+C_e}$; to ensure that the vast majority of forces are positive, it would be enough to take $\overline{p}^2>C_s+C_e$. In 2D systems this implies that
\eq{ \label{eta1}
\eta+g \gtrsim \frac{1}{D^2 \overline{p}^2},
}
so that $\eta+g$ must diverge as the unjamming transition at $\overline{p}=0$ is approached. We recall that both $\eta>0,g>0$ to suppress fluctuations, so in fact both $\eta$ and $g$ must diverge in this limit. 

Manually fixing $\eta+g$ to satisfy \eqref{eta1} is extremely unnatural, because it runs antithetical to the distinction between holographic and `elastic' contributions to the action; $\eta$ and $g$ should not depend on externally controlled quantities, like $\overline{p}$. Instead, a condition like \eqref{eta1} should emerge as a result of imposing $p>0$ in construction of the theory. 
 The deficiency in earlier arguments is apparent: the sampling probability $\omega[\sigmab]$ should impose  $p>0$. Consider a patch of the system with pressure $p(\rv)$. Since the space of force states with all forces repulsive is convex, the volume of force states with pressure $\sim p(\rv)$ will scale as $p(\rv)^{\nu'}$, where $\nu'$ counts the number of force DOF after satisfying local constraints \cite{Tighe08,Tighe11a}. Since we have integrated over all geometric DOF, there will also be a contribution from the number of geometric configurations for a given force configuration. A simple argument suggests that for frictionless disks, the total from mechanical and geometrical contributions gives $\nu' \approx 2$ near the jamming point \footnote{For frictionless disks, there are $N_C = z N/2$ DOF in the forces, where $z$ is the coordination number of the packing. There are additionally $2N$ geometric DOF, the particle centre positions. Force balance removes $2N$ DOF. Near the jamming point, force changes correspond to grain movements of negligible amplitude, so that the DOF can be considered independent. Then the number of free DOF with a fixed mesoscopic pressure is approximated as $N \nu' \approx N_C + 2N-2N-1 \approx z N/2$. At the jamming point, $z=4$, so that $\nu' \approx 2$. A similar argument is made by Henkes and Chakraborty \cite{Henkes09a}.}. Over the entire system, we therefore expect a contribution
\eq{ \label{omp1}
\omega_P[\sigmab] = e^{\nu \int dV \; \log p(\rv)}
}
that enforces positivity of forces, with $\nu = \nu' N/V$. 
 However, if we admit a term $\nu \log p$, then we must admit terms of the form $\p^n (\log p)^q$, with $n \leq d$ by power counting ($\log p$ has zero canonical dimension). Ignoring for simplicity total derivatives, which would correspond to more controllable quantities, we have to add $m |\nabla \log p|^2$. We are thus led to a Lagrangian
\eq{
\Lc_P[\psi] = \Lc[\psi] - \nu \log p[\psi] + \half m |\nabla \log p[\psi]|^2, 
}
which preserves the infinite-dimensional shear symmetry, since the new terms depend only on $p$ \footnote{For any function $f$, the first variation of $\int dV f(p[\psi])$ is $\int dV f'(p) \delta p$. Since, under the shear symmetry, $\delta p=0$, the new terms preserve the symmetry.}. $\Lc_P$ is not Gaussian, and cannot be integrated exactly. The modification of the field equation and the equation of state can be obtained by an expansion to Gaussian order, detailed in Appendix 5. Here we present some results at the classical level, in 2D. First, the leading modification to spatial fluctuations from $\nu$ is
\eq{
C_s = \frac{\Lambda^2}{16 \pi \tilde\eta_R} - C_e
}
where $\tilde\eta_R$ is the renormalized $\tilde\eta=\eta+g$:
 \eq{
 \tilde\eta_R = \tilde\eta + \frac{\nu}{4 \overline{p}^2}, 
 } 
 so that the inequality \eqref{eta1} is then {\it predicted}. This resolves the problem with na\"ive application of the simpler Gaussian model. Simulations of frictionless disk packings indicate that $C_s \propto \pbar^2$ to very good accuracy \cite{Henkes09a,Wu17}, indicating that $\tilde\eta$ plays a sub-dominant role, as will be predicted below. 
 
The field equation becomes nonlinear:
\eq{ \label{class1}
0 = \nabla \nabla : \alphac + \tilde\eta \nabla^4 \psi_c - \nabla^2 \left(\frac{\nu + m \nabla^2 \log \nabla^2 \psi_c }{\nabla^2 \psi_c} \right)
}
with boundary conditions
\eq{ \label{bc4}
0 & = \nv \cdot \left[ \alphac +  \tilde\gamma \sigmab_c +  2 \tilde\eta \; \delb \; p_c - \delb \;\frac{\nu+m \nabla^2 \log p_c}{2p_c} \right], \\
0 & = \nv \cdot \left[ \nabla \cdot \alphac + 2 \tilde\eta \nabla p_c - \nabla \left(\frac{\nu+m\nabla^2 \log p_c}{2p_c}\right) \right] \label{bc5} \\
0 & = \nv \cdot \left[ \nabla \log p_c \right] \label{bc6}
}
From this we read off the equation of state
\eq{ \label{eos3}
0 = \alphac +  \tilde\gamma \sigmabbar +  2 \tilde\eta \; \delb \; \overline{p}  - \delb \; \frac{\nu}{2 \overline{p}}
}
which is easily solved for $\sigmabbar$. Despite its manifest nonlinearity, the classical equation can be solved in some nontrivial limits. Consider a number of sources
\eq{
\alphab = \hat{\overline{\alpha}} + \sum_i \alphab_i \delta(\rv-\rv_i)
}
Then using $ \nabla^2 \log r = 2 \pi \delta(\rv)$ we see that
\eqs{
0 = \nabla^2 \left[ S(\rv) + \tilde\eta \nabla^2 \psi_c - \frac{\nu + m \nabla^2 \log \nabla^2 \psi_c }{\nabla^2 \psi_c} \right]
}
where $S(\rv) = (2 \pi)^{-1} \sum_i \alphac_i : \nabla \nabla \log |\rv-\rv_i|$ {\blue contains the source terms}. This implies
\eq{ \label{c1}
S(\rv) + \tilde\eta \nabla^2 \psi_c - \frac{\nu + m \nabla^2 \log \nabla^2 \psi_c }{\nabla^2 \psi_c} = h(\rv)
}
where $h(\rv)$ is harmonic, $\nabla^2 h = 0$, and fixed by boundary conditions. In asymptotically large systems, we can replace the full boundary conditions \eqref{bc4},\eqref{bc5},\eqref{bc6} with the condition that $\sigmab_c \to \sigmabbar$ as $|\rv| \to \infty$. In this case $h(\rv) \to  h_\infty = 2 \tilde\eta \pbar - \nu/(2 \pbar)$ as $\rv \to \infty$. But then the maximum principle implies that $h$ is constant and equal to $h_\infty$ everywhere.

The parameter $m$ defines a length scale through
\eq{
\xi = \sqrt{\frac{m}{4 \tilde\eta_R \pbar^2}}
}
Let us consider the regime where $\xi \ll 1$ in macroscopic units. Then we can solve \eqref{c1} perturbatively in $\xi$. At $\OO(\xi^0)$, \eqref{c1} is simply a quadratic equation in $\nabla^2 \psi_c$:
\eq{
2 \tilde\eta \nabla^2 \psi_c = h - S(\rv) + \sqrt{(h-S(\rv))^2 + 4 \tilde\eta \nu }.
}
We have taken the positive root to ensure that $p \geq 0$. We can solve the resulting Poisson equation. In an asymptotically large domain,
\eq{
 \psi_c(\rv) = \psi_s(\rv) + \int d^2 r' \; \frac{F\big[h-S(\rvp) \big]}{4 \pi \tilde\eta} \; \log | \rv - \rvp|
}
where $\psi_s(\rv) = \half \rv \times (\sigmabbar-\delb \; \pbar) \times \rv$, which generates the deviatoric part of the mean stress, and $F[h-S]= h - S + \sqrt{(h-S)^2 + 4 \tilde\eta \nu }$. The stress tensor is
\begin{widetext}
\eq{ \label{nonlin4}
\sigmab_c(\rv) = \sigmabbar-\delb \; \pbar + \frac{1}{4\tilde\eta} \delb \; F[h-S(\rv)] + \frac{1}{4\pi \tilde\eta}\int d^2 r' \; F[h-S(\rvp)]  \; \frac{1}{|\rv-\rvp|^2} \left[ \delb - 2 \nv \nv \right]
}
where $\nv = (\rv-\rvp)/|\rv-\rvp|$.

\end{widetext}


At the classical level, the partition function is approximated by its saddle point value, $Z \approx e^{-S_c}$, $S_c = \int \Lc_P[\psi_c]$, and we use \eqref{corr0} to extract correlators. Since the source function $S(\rv)$ appears in a square root in \eqref{nonlin4}, derivatives of $\sigmab_c$ with respect to $\alphab$ will not vanish at any order. This then implies that nontrivial stress correlations will be present at all orders. This should be compared with the Gaussian theory, which predicts only nontrivial pairwise connected correlation functions, with all higher order correlators fixed by Wick's theorem. 


In practice, extracting these higher-order correlators is onerous. Here we will focus on the 2-point pressure correlator, whose computation is outlined in Appendix 5. The result, in large systems and far from the boundary, is
\eq{ \label{pcorr2}
\langle p(\rv) p(0) \rangle_c = A \; \delta(\rv) - (\gamma-g)\frac{(F')^2}{8 \pi \tilde\eta^2} \frac{1}{r^2}, 
}
where $F' = F'(h-S)|_{S=0}=-8\tilde\eta\pbar^2 (4\tilde\eta \pbar^2+\nu)^{-1}$. Here $A$ is the coefficient of the contact terms, which have many contributions. The key result is that the pressure correlator is no longer short-range: it has a power-law tail, whose sign depends on the value of $\gamma-g$. {\blue Since $\gamma$ and $g$ can be independently extracted using the equation of state and by fitting the stress distribution, this result could be directly tested.}

These results can be extended perturbatively in $\xi$. The main effect is to renormalize the function $F$ appearing above: at $\OO(\xi^2)$, it becomes
\eq{
F \to F_R \equiv F + m \frac{\nabla^2 \log F}{\sqrt{(h-S(\rv))^2 + 4 \tilde\eta \nu}}
}
The basic form of correlations is thus not affected, although the coefficients will be. We note that at large $\xi$, a new regime is possible; in fact this regime corresponds to quantum gravity in 2 Euclidean dimensions, for which many results are available \cite{Seiberg90}. This connection will be discussed elsewhere.

\section{Extensions}

{\bf Coulomb friction: } In the previous section we showed that an inequality $p>0$ leads to nontrivial modifications of the theory. For granular materials, and for glasses in general, one may also expect inequalities on the shear stress, such as
\eq{ \label{MC}
\tau \leq \mu p,
}
where $\tau^2 = (\sigma_{ij}-\delta_{ij}p)(\sigma_{ij}-\delta_{ij}p)$ is a deviatoric stress scale, and $\mu$ is a Mohr-Coulomb friction coefficient. In models of granular matter, an inequality of this form is commonly taken to hold exactly at the scale of particle-particle contacts \cite{Andreotti13}. In glasses this is not the case, but at a mesoscopic level, so-called `elastoplastic' models used to investigate plasticity frequently impose a criterion of the form \eqref{MC} \cite{Nicolas17}.  

Similar arguments as applied for the constraint $p>0$ can be applied to \eqref{MC}. One may expect terms of the form $\nu_f \log (\mu^2 p^2 - \tau^2)$, and an associated gradient term. At the Gaussian level, this will renormalize coefficients, while at higher orders one may expect many nontrivial effects, possibly relevant to plasticity. We leave this for future work.



{\bf Structural anisotropy: } We have assumed throughout that the material is isotropic and reflection-symmetric, although it may be subject to external shear stresses, through anisotropic $\alphab$. If subject to continued shear forcing, or if the solid itself has been formed by shear-jamming, then it will retain anisotropy in the particle arrangements, the `fabric.' In this case new terms should be added to the Lagrangian. The lowest-order such term is $g_a F_{ij} \sigma_{ij} p$, where $F_{ij}$ is a traceless fabric tensor. It is straightforward to extend the previous results to this case; for the Gaussian model, the field equation will resemble that derived by the author in previous work \cite{DeGiuli14a}. The fluctuations will be modified from the isotropic case, and long-range correlations will be present even in the pressure correlator. We leave a full discussion of these effects for future work.

{\blue Note also that gravity is a particular case of anisotropy, which we have also neglected. A net force such as gravity cannot be accommodated by stress functions, which identically satisfy $\nabla \cdot \sigmab = 0$. Thus in the presence of gravity, our theory describes the stress state in the subspace satisfying $\nabla \cdot \sigmab = 0$. }

{\bf Topological excitations: } Since stresses are given, in the gauge formulation, by curvature of 
 stress functions, these functions are not {\it a priori} required to be single-valued. If not, these could be subject to vortex-like topological excitations, familiar from the theory of 2D melting \cite{Strandburg88,Beekman17}. In fact, it was shown in \cite{DeGiuli14a} by explicit construction at the particle scale that the 2D discrete Airy stress function is continuous at the smallest scale at which it can be defined, thus precluding such excitations in 2D. It is not known whether this result survives in 3D.

{\blue
{\bf Geometrical variables: } We have restricted our theory to the stress ensemble, where any geometrical variables are assumed to have been marginalized over. Previous work has considered the volume as an important additional holographic quantity \cite{Edwards89,Edwards05,Makse02,Metzger04,Blumenfeld12,Asenjo14}. At the simplest level of description, a complete theory will lead to a temperature-like coupling for the volume, the `compactivity', and volume-stress couplings. If the volume only appears up to quadratic order, then it can be integrated out, and the reduced stress ensemble derived explicitly. In this case we will obtain the stress ensemble considered here, with coupling constants renormalized from their bare values. If, however, nontrivial constraints are added to the theory, such as non-penetration of hard grains, then a non-Gaussian coupled theory will result. We leave consideration of this for future work.
}

\section{Discussion}
\subsection{Behavior near the jamming point}

{\blue Much of the recent work on amorphous solids has focussed on their behavior close to the jamming point where solids lose rigidity altogether. A system frequently used in numerical simulations is a packing of soft frictionless spheres, with a one-sided linear elastic contact interaction, so that forces are always repulsive \cite{OHern03}. 

We have argued above that the `elastic' parameters $\eta$ and $g$ should not depend on the external driving conditions, including the pressure. Dimensionally, $\eta$ and $g$ both have units of length$^d/$stress$^2$. For this model, they must then be proportional to $1/k^2$, where $k$ is the spring constant of the contact interaction. At the jamming point $\pbar/(k D^{2-d}) \to 0$, thus $\eta \pbar^2 \ll \nu$ and some simplifications occur. For example, $\tilde \eta_R \approx \nu/(4\pbar^2)$, explaining the scaling of stress fluctuations observed in \cite{Henkes09a,Wu17}. Also, the term involving $\tilde \eta$ can be dropped in Eq.\eqref{eos3}.
}

\subsection{Comparison with data}

{\bf Equation of state:} Numerous works in the granular matter community have attempted to test proposed forms of $\PP[\sigmab]$ and extract the temperature-like quantities $\alphab$ and $\gamma$ from numerical and experimental data \cite{Henkes07,Henkes09a,Puckett13,Wu15a}. To explain the technique, let us consider the theory with generic sampling probability $\omega$ and write $\PP[\sigmab]$ as
\eq{ \label{Psig2}
\PP[\sigmab] = \frac{\omega[\sigmab]}{Z(\alphab,\gamma)} e^{-\alphab : \hat{\Gamma} - \gamma \Ac}
}
where $\hat{\Gamma} = V \overline{\sigmab}$ and $\Ac = V \overline{\det \sigma}$ over a subsystem. $\alphab$ and $\gamma$ are set by the global value of $\sigmab$ through the equation of state. We can then consider this equation as expressing the conditional probability of observing $\hat{\Gamma}$ and $\Ac$ in a subsystem with fixed $\alphab,\gamma$, i.e.
\eq{ \label{Psig3}
\PP[\hat{\Gamma},\Ac | \alphab,\gamma] = \frac{\omega[\hat{\Gamma},\Ac | \alphab,\gamma]}{Z(\alphab,\gamma)} e^{-\alphab : \hat{\Gamma} - \gamma \Ac},
}
where there is an implied integration needed to produce the density of states depending only on $\Gammab$ and $\Ac$. Evaluating this at two pairs of values of $\alphab,\gamma$, but the same values of $\hat{\Gamma},\Ac$, one finds
\eq{
R & \equiv \log \frac{\PP[\hat{\Gamma},\Ac | \alphab_1,\gamma_1]}{\PP[\hat{\Gamma},\Ac | \alphab_2,\gamma_2]} \notag \\
& = \log \frac{\omega_1 Z_2}{\omega_2 Z_1} - (\alphab_1 - \alphab_2) : \hat{\Gamma} - (\gamma_1-\gamma_2) \Ac 
}
where $\omega_i = \omega[\hat{\Gamma},\Ac | \alphab_i,\gamma_i], \; Z_i = Z(\alphab_i,\gamma_i)$. The goal of the `overlapping histograms' method is to use the linear dependence of the latter terms in $R$ on $\Gammab$ and $\Ac$ to extract $\alphab$ and $\gamma$ up to additive constants. Of course, this is only possible if the dependence of $\omega_2/\omega_1$ on $\hat{\Gamma}$ and $\Ac$ can be neglected. We have argued above that the parameters appearing in the sampling probability, such as $\eta$ and $g$, should not depend on external parameters like the pressure. As discussed above, to reconcile this fact with the observed behavior that suggests $\tilde\eta_R \propto 1/\pbar^2$, one needs to incorporate nontrivial $\log p$ terms in the sampling probability. After this modification, the assumption that $\omega$ is indeed independent of globally applied stresses is theoretically and empirically justified. Thus under such an assumption, $\omega_i = \omega[\Gammab,\Ac]$, and these terms cancel from $R$. 

Let us note that since $\Ac = \int dV \det \sigmab$ in the continuum, in isotropic packings we expect $\Ac \approx V p^2$ and $\Gammab \approx \delb V p$. Then \eqref{Psig2} implies a non-centered Gaussian distribution for the pressure, with modifications from $\omega$.

The first works to apply this method to the stress ensemble \cite{Henkes07,Henkes09a} considered isotropic packings of frictionless disks, and the Maxwell-Cremona area $\Ac$ was not considered. The authors used the difference $R(\Gamma_1)-R(\Gamma_2)$ and found that $\alpha \propto 1/p$. This is recovered in our theory from Eq.\eqref{eos3} if $\eta\approx\gamma\approx g\approx 0$, and indeed $\gamma$ was not considered in the corresponding theory \cite{Henkes09a}. Experimental results on a system of nearly frictionless photoelastic disks found consistent results \cite{Puckett13}.

More recently, a very careful analysis of simulations of the same system \cite{Wu15a} showed that the method applied in \cite{Henkes07,Henkes09a} can lead to false positives in fitting of \eqref{Psig2} with $\gamma=0$. Their refined analysis using $R$ concluded that $\gamma$ needs to be included in the analysis. The resulting measurements found that, approximately, $\alpha \propto -1/p$ and $\gamma \propto 1/p^2$, consistent with our \eqref{eos3}.

Recently, experimental results were analyzed with the full ensemble \eqref{Psig2}, and measured both $\alpha$ and $\gamma$ \cite{Bililign18}. It was found that $\alpha$ can depend on the experimental protocol, consistent with its identification in our theory as conjugate to a controllable variable. 

{\blue In all of these works, the systems considered had repulsive-only forces, and were close to the unjamming point. The natural equation-of-state for such materials is Eq. \eqref{eos3}, where $\tilde \eta$ can be set to zero as described in the previous section. In its isotropic version $\alphab = \alpha \delb$ the equation reads $\alpha + \gamma \pbar = \nu/(2\pbar)$. This should be compared with the isotropic equation-of-state in the Gaussian theory, Eq.\ref{eos1}, which reads $\alpha +  \gamma \pbar = -\pbar (g + 2\eta)$. Since the two equations have the same left-hand side, but right-hand sides with differing sign, these could be used to discriminate between the Gaussian and non-Gaussian theories. Note that in Ref. \cite{Henkes09a}, the authors derived an equation-of-state from microscopic considerations, of the form $\alpha = a/\pbar$, consistent with our result from the non-Gaussian theory when $\gamma=0$. These authors did not derive an equation-of-state from their field theory, but the result must be the same as ours for the Gaussian theory, with $\gamma=0$ and isotropic $\alphab$, since the two theories coincide in this particular case. }

{\bf Anomalous Stress correlations: } To our knowledge, the large body of experimental and numerical work on stress correlators is compatible with the simple Gaussian theory, with one exception. This exception is Ref.\onlinecite{Wu17}, where it was found that packings of frictionless particles and Lennard-Jones glasses, both in 2D, have anomalously large stress correlations at small wavenumber $k \lesssim 0.1 D^{-1}$, with approximately $\langle |p(k)|^2 \rangle_c \sim k^{-1.3}$, and similar results for other stress correlators.

We have shown above that when repulsive constraints are considered, the pressure correlator has long-range correlations, Eq. \eqref{pcorr2}. In Fourier space, these will take the form
\eq{
\langle |p(k)|^2 \rangle_c = A - C \int \frac{dr}{r} J_0(kr),
}
where $J_0$ is a Bessel function. The integral has a logarithmic divergence at small $r$, and must be cutoff with a length scale $\ell$. We find then that $\langle |p(k)|^2 \rangle_c - A \propto (\ell k)^{-3/2}$ for $k \ell \gg 1$, while $\langle |p(k)|^2 \rangle_c - A \propto \log k$ for $k \ell \lesssim 1$. Thus this result is consistent with the measurements of \cite{Wu17}, but only if the length scale $\ell$ is very large. Within the present theory, there is no particular reason to expect a large $\ell$, but we cannot exclude this possibility, so we leave this result as a tentative, but promising, explanation of the anomaly reported in \cite{Wu17}.


\subsection{Conclusion} 
We have theoretically shown, in both 2D and 3D, that athermal amorphous solids have long-range stress correlations. Explicit forms of these correlations were derived in a field theory that is applicable at long length scales. The main assumptions underlying the theory are that: (i) all quantities are probed at lengths much larger than the particle size; (ii) all interactions between the stresses are themselves local
; and (iii) the material is isotropic. Furthermore, we derived the equations of state relating the magnitude of fluctuations to imposed stresses, and field equations that can be used to find the spatial form of stresses in arbitrary domains. We also identified a new holographic quantity in 3D systems. 

The predicted form of stress correlators is in extremely good agreement with simulations on supercooled liquids \cite{Lemaitre14,Lemaitre15,Lemaitre17}. Besides the basic functional form and anisotropy, we have been able to explain several minute features of the correlator, going beyond what is possible from a strictly dynamical, elastic theory \cite{Lemaitre14}. {\blue This supports the claim that the structure of inherent states, as characterized by their stress, follows from considerations of mechanical equilibrium alone, as we have argued. }

For athermal amorphous solids dominated by repulsive interactions, the Gaussian theory leads to a paradox, which we resolved. The requirement that pressure $p$ remains positive leads to a non-Gaussian theory, with new features: the equation of state is modified, and nontrivial stress correlations are predicted at all orders. The equation of state agrees with previous tests of the ensemble. We find that the pressure has long-range correlations, which may explain anomalous stress correlations found in \cite{Wu17}. 

We identified an infinite-dimensional symmetry, corresponding to increments of shear stress which do not affect the action except through boundary terms. {\blue If the system is in some local maximum of probability in configuration space, equivalent to a local energy minimum in a thermal system, then the most probable route out of this state to another inherent state will be through the action of this symmetry. Thus transitions from one inherent state to another will largely proceed by changes in the deviatoric stress. } This symmetry is expected to be very important for plasticity dynamics, to be considered in the future.

Our analysis has been restricted to the saddle-point level, which is exact for the Gaussian theory, but not for the $p>0$ theory. There are five sources which may lead to renormalization of the discussed long-range correlations: (i) the terms arising from holographic quantities that live on the boundary will lead to nontrivial boundary effects, which we have not investigated; (ii) for the $p>0$ theory, fluctuations will modify the pressure correlator; (iii) coupling to geometric variables can lead to nontrivial pressure fluctuations if the effective coupling after integrating out geometric variables is nonlocal; (iv) geometric variables may admit topological excitations, like contact-opening excitations \cite{DeGiuli15a}; and (v) structural anisotropy will lead to shear-like effects in the pressure correlator. It would be valuable to pursue these many directions of future research.

Overall, our description of the inherent states of athermal amorphous solids is simple and widely applicable. A crucial application is to understand the effect of long-range stress correlations on vibrational properties \cite{Gelin16}; this will be tackled in future work.



\begin{acknowledgments} The author kindly acknowledges many exchanges within the Simons `Cracking the glass' collaboration, especially frequent discussions with E. Lerner.
\end{acknowledgments}

\bibliography{../Glasses} 

\begin{thebibliography}{84}%
\makeatletter
\providecommand \@ifxundefined [1]{%
 \@ifx{#1\undefined}
}%
\providecommand \@ifnum [1]{%
 \ifnum #1\expandafter \@firstoftwo
 \else \expandafter \@secondoftwo
 \fi
}%
\providecommand \@ifx [1]{%
 \ifx #1\expandafter \@firstoftwo
 \else \expandafter \@secondoftwo
 \fi
}%
\providecommand \natexlab [1]{#1}%
\providecommand \enquote  [1]{``#1''}%
\providecommand \bibnamefont  [1]{#1}%
\providecommand \bibfnamefont [1]{#1}%
\providecommand \citenamefont [1]{#1}%
\providecommand \href@noop [0]{\@secondoftwo}%
\providecommand \href [0]{\begingroup \@sanitize@url \@href}%
\providecommand \@href[1]{\@@startlink{#1}\@@href}%
\providecommand \@@href[1]{\endgroup#1\@@endlink}%
\providecommand \@sanitize@url [0]{\catcode `\\12\catcode `\$12\catcode
  `\&12\catcode `\#12\catcode `\^12\catcode `\_12\catcode `\%12\relax}%
\providecommand \@@startlink[1]{}%
\providecommand \@@endlink[0]{}%
\providecommand \url  [0]{\begingroup\@sanitize@url \@url }%
\providecommand \@url [1]{\endgroup\@href {#1}{\urlprefix }}%
\providecommand \urlprefix  [0]{URL }%
\providecommand \Eprint [0]{\href }%
\providecommand \doibase [0]{http://dx.doi.org/}%
\providecommand \selectlanguage [0]{\@gobble}%
\providecommand \bibinfo  [0]{\@secondoftwo}%
\providecommand \bibfield  [0]{\@secondoftwo}%
\providecommand \translation [1]{[#1]}%
\providecommand \BibitemOpen [0]{}%
\providecommand \bibitemStop [0]{}%
\providecommand \bibitemNoStop [0]{.\EOS\space}%
\providecommand \EOS [0]{\spacefactor3000\relax}%
\providecommand \BibitemShut  [1]{\csname bibitem#1\endcsname}%
\let\auto@bib@innerbib\@empty
\bibitem [{\citenamefont {Henkes}\ and\ \citenamefont
  {Chakraborty}(2009)}]{Henkes09a}%
  \BibitemOpen
  \bibfield  {author} {\bibinfo {author} {\bibfnamefont {S.}~\bibnamefont
  {Henkes}}\ and\ \bibinfo {author} {\bibfnamefont {B.}~\bibnamefont
  {Chakraborty}},\ }\href {\doibase 10.1103/PhysRevE.79.061301} {\bibfield
  {journal} {\bibinfo  {journal} {Phys. Rev. E}\ }\textbf {\bibinfo {volume}
  {79}},\ \bibinfo {pages} {061301} (\bibinfo {year} {2009})}\BibitemShut
  {NoStop}%
\bibitem [{\citenamefont {Wu}\ \emph {et~al.}(2017)\citenamefont {Wu},
  \citenamefont {Karimi}, \citenamefont {Maloney},\ and\ \citenamefont
  {Teitel}}]{Wu17}%
  \BibitemOpen
  \bibfield  {author} {\bibinfo {author} {\bibfnamefont {Y.}~\bibnamefont
  {Wu}}, \bibinfo {author} {\bibfnamefont {K.}~\bibnamefont {Karimi}}, \bibinfo
  {author} {\bibfnamefont {C.~E.}\ \bibnamefont {Maloney}}, \ and\ \bibinfo
  {author} {\bibfnamefont {S.}~\bibnamefont {Teitel}},\ }\href@noop {}
  {\bibfield  {journal} {\bibinfo  {journal} {Physical Review E}\ }\textbf
  {\bibinfo {volume} {96}},\ \bibinfo {pages} {032902} (\bibinfo {year}
  {2017})}\BibitemShut {NoStop}%
\bibitem [{\citenamefont {Lema{\^\i}tre}(2014)}]{Lemaitre14}%
  \BibitemOpen
  \bibfield  {author} {\bibinfo {author} {\bibfnamefont {A.}~\bibnamefont
  {Lema{\^\i}tre}},\ }\href@noop {} {\bibfield  {journal} {\bibinfo  {journal}
  {Physical review letters}\ }\textbf {\bibinfo {volume} {113}},\ \bibinfo
  {pages} {245702} (\bibinfo {year} {2014})}\BibitemShut {NoStop}%
\bibitem [{\citenamefont {Wu}\ \emph {et~al.}(2015)\citenamefont {Wu},
  \citenamefont {Iwashita},\ and\ \citenamefont {Egami}}]{Wu15}%
  \BibitemOpen
  \bibfield  {author} {\bibinfo {author} {\bibfnamefont {B.}~\bibnamefont
  {Wu}}, \bibinfo {author} {\bibfnamefont {T.}~\bibnamefont {Iwashita}}, \ and\
  \bibinfo {author} {\bibfnamefont {T.}~\bibnamefont {Egami}},\ }\href@noop {}
  {\bibfield  {journal} {\bibinfo  {journal} {Physical Review E}\ }\textbf
  {\bibinfo {volume} {91}},\ \bibinfo {pages} {032301} (\bibinfo {year}
  {2015})}\BibitemShut {NoStop}%
\bibitem [{\citenamefont {Lema{\^\i}tre}(2015)}]{Lemaitre15}%
  \BibitemOpen
  \bibfield  {author} {\bibinfo {author} {\bibfnamefont {A.}~\bibnamefont
  {Lema{\^\i}tre}},\ }\href@noop {} {\bibfield  {journal} {\bibinfo  {journal}
  {The Journal of chemical physics}\ }\textbf {\bibinfo {volume} {143}},\
  \bibinfo {pages} {164515} (\bibinfo {year} {2015})}\BibitemShut {NoStop}%
\bibitem [{\citenamefont {Chikkadi}\ \emph {et~al.}(2011)\citenamefont
  {Chikkadi}, \citenamefont {Wegdam}, \citenamefont {Bonn}, \citenamefont
  {Nienhuis},\ and\ \citenamefont {Schall}}]{Chikkadi11}%
  \BibitemOpen
  \bibfield  {author} {\bibinfo {author} {\bibfnamefont {V.}~\bibnamefont
  {Chikkadi}}, \bibinfo {author} {\bibfnamefont {G.}~\bibnamefont {Wegdam}},
  \bibinfo {author} {\bibfnamefont {D.}~\bibnamefont {Bonn}}, \bibinfo {author}
  {\bibfnamefont {B.}~\bibnamefont {Nienhuis}}, \ and\ \bibinfo {author}
  {\bibfnamefont {P.}~\bibnamefont {Schall}},\ }\href@noop {} {\bibfield
  {journal} {\bibinfo  {journal} {Physical Review Letters}\ }\textbf {\bibinfo
  {volume} {107}},\ \bibinfo {pages} {198303} (\bibinfo {year}
  {2011})}\BibitemShut {NoStop}%
\bibitem [{\citenamefont {Jensen}\ \emph {et~al.}(2014)\citenamefont {Jensen},
  \citenamefont {Weitz},\ and\ \citenamefont {Spaepen}}]{Jensen14}%
  \BibitemOpen
  \bibfield  {author} {\bibinfo {author} {\bibfnamefont {K.}~\bibnamefont
  {Jensen}}, \bibinfo {author} {\bibfnamefont {D.~A.}\ \bibnamefont {Weitz}}, \
  and\ \bibinfo {author} {\bibfnamefont {F.}~\bibnamefont {Spaepen}},\
  }\href@noop {} {\bibfield  {journal} {\bibinfo  {journal} {Physical Review
  E}\ }\textbf {\bibinfo {volume} {90}},\ \bibinfo {pages} {042305} (\bibinfo
  {year} {2014})}\BibitemShut {NoStop}%
\bibitem [{\citenamefont {Illing}\ \emph {et~al.}(2016)\citenamefont {Illing},
  \citenamefont {Fritschi}, \citenamefont {Hajnal}, \citenamefont {Klix},
  \citenamefont {Keim},\ and\ \citenamefont {Fuchs}}]{Illing16}%
  \BibitemOpen
  \bibfield  {author} {\bibinfo {author} {\bibfnamefont {B.}~\bibnamefont
  {Illing}}, \bibinfo {author} {\bibfnamefont {S.}~\bibnamefont {Fritschi}},
  \bibinfo {author} {\bibfnamefont {D.}~\bibnamefont {Hajnal}}, \bibinfo
  {author} {\bibfnamefont {C.}~\bibnamefont {Klix}}, \bibinfo {author}
  {\bibfnamefont {P.}~\bibnamefont {Keim}}, \ and\ \bibinfo {author}
  {\bibfnamefont {M.}~\bibnamefont {Fuchs}},\ }\href@noop {} {\bibfield
  {journal} {\bibinfo  {journal} {Physical review letters}\ }\textbf {\bibinfo
  {volume} {117}},\ \bibinfo {pages} {208002} (\bibinfo {year}
  {2016})}\BibitemShut {NoStop}%
\bibitem [{\citenamefont {Le~Bouil}\ \emph {et~al.}(2014)\citenamefont
  {Le~Bouil}, \citenamefont {Amon}, \citenamefont {McNamara},\ and\
  \citenamefont {Crassous}}]{Le-Bouil14}%
  \BibitemOpen
  \bibfield  {author} {\bibinfo {author} {\bibfnamefont {A.}~\bibnamefont
  {Le~Bouil}}, \bibinfo {author} {\bibfnamefont {A.}~\bibnamefont {Amon}},
  \bibinfo {author} {\bibfnamefont {S.}~\bibnamefont {McNamara}}, \ and\
  \bibinfo {author} {\bibfnamefont {J.}~\bibnamefont {Crassous}},\ }\href@noop
  {} {\bibfield  {journal} {\bibinfo  {journal} {Physical review letters}\
  }\textbf {\bibinfo {volume} {112}},\ \bibinfo {pages} {246001} (\bibinfo
  {year} {2014})}\BibitemShut {NoStop}%
\bibitem [{\citenamefont {Bi}\ \emph {et~al.}(2013)\citenamefont {Bi},
  \citenamefont {Zhang}, \citenamefont {Behringer},\ and\ \citenamefont
  {Chakraborty}}]{Bi13}%
  \BibitemOpen
  \bibfield  {author} {\bibinfo {author} {\bibfnamefont {D.}~\bibnamefont
  {Bi}}, \bibinfo {author} {\bibfnamefont {J.}~\bibnamefont {Zhang}}, \bibinfo
  {author} {\bibfnamefont {R.}~\bibnamefont {Behringer}}, \ and\ \bibinfo
  {author} {\bibfnamefont {B.}~\bibnamefont {Chakraborty}},\ }\href@noop {}
  {\bibfield  {journal} {\bibinfo  {journal} {EPL (Europhysics Letters)}\
  }\textbf {\bibinfo {volume} {102}},\ \bibinfo {pages} {34002} (\bibinfo
  {year} {2013})}\BibitemShut {NoStop}%
\bibitem [{\citenamefont {Sarkar}\ \emph {et~al.}(2013)\citenamefont {Sarkar},
  \citenamefont {Bi}, \citenamefont {Zhang}, \citenamefont {Behringer},\ and\
  \citenamefont {Chakraborty}}]{Sarkar13}%
  \BibitemOpen
  \bibfield  {author} {\bibinfo {author} {\bibfnamefont {S.}~\bibnamefont
  {Sarkar}}, \bibinfo {author} {\bibfnamefont {D.}~\bibnamefont {Bi}}, \bibinfo
  {author} {\bibfnamefont {J.}~\bibnamefont {Zhang}}, \bibinfo {author}
  {\bibfnamefont {R.}~\bibnamefont {Behringer}}, \ and\ \bibinfo {author}
  {\bibfnamefont {B.}~\bibnamefont {Chakraborty}},\ }\href@noop {} {\bibfield
  {journal} {\bibinfo  {journal} {Physical review letters}\ }\textbf {\bibinfo
  {volume} {111}},\ \bibinfo {pages} {068301} (\bibinfo {year}
  {2013})}\BibitemShut {NoStop}%
\bibitem [{\citenamefont {Karmakar}\ \emph {et~al.}(2010)\citenamefont
  {Karmakar}, \citenamefont {Lerner},\ and\ \citenamefont
  {Procaccia}}]{Karmakar10a}%
  \BibitemOpen
  \bibfield  {author} {\bibinfo {author} {\bibfnamefont {S.}~\bibnamefont
  {Karmakar}}, \bibinfo {author} {\bibfnamefont {E.}~\bibnamefont {Lerner}}, \
  and\ \bibinfo {author} {\bibfnamefont {I.}~\bibnamefont {Procaccia}},\ }\href
  {\doibase 10.1103/PhysRevE.82.055103} {\bibfield  {journal} {\bibinfo
  {journal} {Phys. Rev. E}\ }\textbf {\bibinfo {volume} {82}},\ \bibinfo
  {pages} {055103} (\bibinfo {year} {2010})}\BibitemShut {NoStop}%
\bibitem [{\citenamefont {Lema\^itre}\ and\ \citenamefont
  {Caroli}(2009)}]{Lemaitre09}%
  \BibitemOpen
  \bibfield  {author} {\bibinfo {author} {\bibfnamefont {A.}~\bibnamefont
  {Lema\^itre}}\ and\ \bibinfo {author} {\bibfnamefont {C.}~\bibnamefont
  {Caroli}},\ }\href {\doibase 10.1103/PhysRevLett.103.065501} {\bibfield
  {journal} {\bibinfo  {journal} {Phys. Rev. Lett.}\ }\textbf {\bibinfo
  {volume} {103}},\ \bibinfo {pages} {065501} (\bibinfo {year}
  {2009})}\BibitemShut {NoStop}%
\bibitem [{\citenamefont {Maloney}\ and\ \citenamefont
  {Lemaitre}(2004)}]{Maloney04}%
  \BibitemOpen
  \bibfield  {author} {\bibinfo {author} {\bibfnamefont {C.}~\bibnamefont
  {Maloney}}\ and\ \bibinfo {author} {\bibfnamefont {A.}~\bibnamefont
  {Lemaitre}},\ }\href {\doibase 10.1103/PhysRevLett.93.016001} {\bibfield
  {journal} {\bibinfo  {journal} {Phys. Rev. Lett.}\ }\textbf {\bibinfo
  {volume} {93}},\ \bibinfo {pages} {016001} (\bibinfo {year}
  {2004})}\BibitemShut {NoStop}%
\bibitem [{\citenamefont {Maloney}\ and\ \citenamefont
  {Lema\^itre}(2006)}]{Maloney06a}%
  \BibitemOpen
  \bibfield  {author} {\bibinfo {author} {\bibfnamefont {C.~E.}\ \bibnamefont
  {Maloney}}\ and\ \bibinfo {author} {\bibfnamefont {A.}~\bibnamefont
  {Lema\^itre}},\ }\href@noop {} {\bibfield  {journal} {\bibinfo  {journal}
  {Phys. Rev. E}\ }\textbf {\bibinfo {volume} {74}},\ \bibinfo {pages} {016118}
  (\bibinfo {year} {2006})}\BibitemShut {NoStop}%
\bibitem [{\citenamefont {Amon}\ \emph {et~al.}(2012)\citenamefont {Amon},
  \citenamefont {Nguyen}, \citenamefont {Bruand}, \citenamefont {Crassous},\
  and\ \citenamefont {Cl\'ement}}]{Amon2012}%
  \BibitemOpen
  \bibfield  {author} {\bibinfo {author} {\bibfnamefont {A.}~\bibnamefont
  {Amon}}, \bibinfo {author} {\bibfnamefont {V.~B.}\ \bibnamefont {Nguyen}},
  \bibinfo {author} {\bibfnamefont {A.}~\bibnamefont {Bruand}}, \bibinfo
  {author} {\bibfnamefont {J.}~\bibnamefont {Crassous}}, \ and\ \bibinfo
  {author} {\bibfnamefont {E.}~\bibnamefont {Cl\'ement}},\ }\href {\doibase
  10.1103/PhysRevLett.108.135502} {\bibfield  {journal} {\bibinfo  {journal}
  {Phys. Rev. Lett.}\ }\textbf {\bibinfo {volume} {108}},\ \bibinfo {pages}
  {135502} (\bibinfo {year} {2012})}\BibitemShut {NoStop}%
\bibitem [{\citenamefont {H{\'e}braud}\ \emph {et~al.}(1997)\citenamefont
  {H{\'e}braud}, \citenamefont {Lequeux}, \citenamefont {Munch},\ and\
  \citenamefont {Pine}}]{Hebraud97}%
  \BibitemOpen
  \bibfield  {author} {\bibinfo {author} {\bibfnamefont {P.}~\bibnamefont
  {H{\'e}braud}}, \bibinfo {author} {\bibfnamefont {F.}~\bibnamefont
  {Lequeux}}, \bibinfo {author} {\bibfnamefont {J.}~\bibnamefont {Munch}}, \
  and\ \bibinfo {author} {\bibfnamefont {D.}~\bibnamefont {Pine}},\ }\href@noop
  {} {\bibfield  {journal} {\bibinfo  {journal} {Physical Review Letters}\
  }\textbf {\bibinfo {volume} {78}},\ \bibinfo {pages} {4657} (\bibinfo {year}
  {1997})}\BibitemShut {NoStop}%
\bibitem [{\citenamefont {Gartner}\ and\ \citenamefont
  {Lerner}(2016)}]{Gartner16}%
  \BibitemOpen
  \bibfield  {author} {\bibinfo {author} {\bibfnamefont {L.}~\bibnamefont
  {Gartner}}\ and\ \bibinfo {author} {\bibfnamefont {E.}~\bibnamefont
  {Lerner}},\ }\href@noop {} {\bibfield  {journal} {\bibinfo  {journal}
  {Physical Review E}\ }\textbf {\bibinfo {volume} {93}},\ \bibinfo {pages}
  {011001} (\bibinfo {year} {2016})}\BibitemShut {NoStop}%
\bibitem [{\citenamefont {Zylberg}\ \emph {et~al.}(2017)\citenamefont
  {Zylberg}, \citenamefont {Lerner}, \citenamefont {Bar-Sinai},\ and\
  \citenamefont {Bouchbinder}}]{Zylberg17}%
  \BibitemOpen
  \bibfield  {author} {\bibinfo {author} {\bibfnamefont {J.}~\bibnamefont
  {Zylberg}}, \bibinfo {author} {\bibfnamefont {E.}~\bibnamefont {Lerner}},
  \bibinfo {author} {\bibfnamefont {Y.}~\bibnamefont {Bar-Sinai}}, \ and\
  \bibinfo {author} {\bibfnamefont {E.}~\bibnamefont {Bouchbinder}},\
  }\href@noop {} {\bibfield  {journal} {\bibinfo  {journal} {Proceedings of the
  National Academy of Sciences}\ }\textbf {\bibinfo {volume} {114}},\ \bibinfo
  {pages} {7289} (\bibinfo {year} {2017})}\BibitemShut {NoStop}%
\bibitem [{\citenamefont {Lin}\ \emph {et~al.}(2014{\natexlab{a}})\citenamefont
  {Lin}, \citenamefont {Lerner}, \citenamefont {Rosso},\ and\ \citenamefont
  {Wyart}}]{Lin14}%
  \BibitemOpen
  \bibfield  {author} {\bibinfo {author} {\bibfnamefont {J.}~\bibnamefont
  {Lin}}, \bibinfo {author} {\bibfnamefont {E.}~\bibnamefont {Lerner}},
  \bibinfo {author} {\bibfnamefont {A.}~\bibnamefont {Rosso}}, \ and\ \bibinfo
  {author} {\bibfnamefont {M.}~\bibnamefont {Wyart}},\ }\href@noop {}
  {\bibfield  {journal} {\bibinfo  {journal} {Proceedings of the National
  Academy of Sciences}\ }\textbf {\bibinfo {volume} {111}},\ \bibinfo {pages}
  {14382} (\bibinfo {year} {2014}{\natexlab{a}})}\BibitemShut {NoStop}%
\bibitem [{\citenamefont {Lin}\ \emph {et~al.}(2014{\natexlab{b}})\citenamefont
  {Lin}, \citenamefont {Saade}, \citenamefont {Lerner}, \citenamefont {Rosso},\
  and\ \citenamefont {Wyart}}]{Lin14a}%
  \BibitemOpen
  \bibfield  {author} {\bibinfo {author} {\bibfnamefont {J.}~\bibnamefont
  {Lin}}, \bibinfo {author} {\bibfnamefont {A.}~\bibnamefont {Saade}}, \bibinfo
  {author} {\bibfnamefont {E.}~\bibnamefont {Lerner}}, \bibinfo {author}
  {\bibfnamefont {A.}~\bibnamefont {Rosso}}, \ and\ \bibinfo {author}
  {\bibfnamefont {M.}~\bibnamefont {Wyart}},\ }\href@noop {} {\bibfield
  {journal} {\bibinfo  {journal} {EPL (Europhysics Letters)}\ }\textbf
  {\bibinfo {volume} {105}},\ \bibinfo {pages} {26003} (\bibinfo {year}
  {2014}{\natexlab{b}})}\BibitemShut {NoStop}%
\bibitem [{\citenamefont {Nicolas}\ \emph {et~al.}(2017)\citenamefont
  {Nicolas}, \citenamefont {Ferrero}, \citenamefont {Martens},\ and\
  \citenamefont {Barrat}}]{Nicolas17}%
  \BibitemOpen
  \bibfield  {author} {\bibinfo {author} {\bibfnamefont {A.}~\bibnamefont
  {Nicolas}}, \bibinfo {author} {\bibfnamefont {E.~E.}\ \bibnamefont
  {Ferrero}}, \bibinfo {author} {\bibfnamefont {K.}~\bibnamefont {Martens}}, \
  and\ \bibinfo {author} {\bibfnamefont {J.-L.}\ \bibnamefont {Barrat}},\
  }\href@noop {} {\bibfield  {journal} {\bibinfo  {journal} {arXiv preprint
  arXiv:1708.09194}\ } (\bibinfo {year} {2017})}\BibitemShut {NoStop}%
\bibitem [{\citenamefont {Roux}\ and\ \citenamefont {Combe}(2010)}]{Roux10}%
  \BibitemOpen
  \bibfield  {author} {\bibinfo {author} {\bibfnamefont {J.}~\bibnamefont
  {Roux}}\ and\ \bibinfo {author} {\bibfnamefont {G.}~\bibnamefont {Combe}},\
  }in\ \href@noop {} {\emph {\bibinfo {booktitle} {Proceedings of the
  IUTAM-ISIMM Symposium on Mathematical Modeling and Physical Instances of
  Granular Flow}}},\ Vol.\ \bibinfo {volume} {1227}\ (\bibinfo  {publisher}
  {AIP},\ \bibinfo {year} {2010})\ pp.\ \bibinfo {pages} {260--270}\BibitemShut
  {NoStop}%
\bibitem [{\citenamefont {Schreck}\ \emph {et~al.}(2011)\citenamefont
  {Schreck}, \citenamefont {Bertrand}, \citenamefont {O'Hern},\ and\
  \citenamefont {Shattuck}}]{Schreck11}%
  \BibitemOpen
  \bibfield  {author} {\bibinfo {author} {\bibfnamefont {C.~F.}\ \bibnamefont
  {Schreck}}, \bibinfo {author} {\bibfnamefont {T.}~\bibnamefont {Bertrand}},
  \bibinfo {author} {\bibfnamefont {C.~S.}\ \bibnamefont {O'Hern}}, \ and\
  \bibinfo {author} {\bibfnamefont {M.~D.}\ \bibnamefont {Shattuck}},\ }\href
  {\doibase 10.1103/PhysRevLett.107.078301} {\bibfield  {journal} {\bibinfo
  {journal} {Phys. Rev. Lett.}\ }\textbf {\bibinfo {volume} {107}},\ \bibinfo
  {pages} {078301} (\bibinfo {year} {2011})}\BibitemShut {NoStop}%
\bibitem [{\citenamefont {Lema{\^\i}tre}(2017)}]{Lemaitre17}%
  \BibitemOpen
  \bibfield  {author} {\bibinfo {author} {\bibfnamefont {A.}~\bibnamefont
  {Lema{\^\i}tre}},\ }\href@noop {} {\bibfield  {journal} {\bibinfo  {journal}
  {Physical Review E}\ }\textbf {\bibinfo {volume} {96}},\ \bibinfo {pages}
  {052101} (\bibinfo {year} {2017})}\BibitemShut {NoStop}%
\bibitem [{\citenamefont {Edwards}(2005)}]{Edwards05}%
  \BibitemOpen
  \bibfield  {author} {\bibinfo {author} {\bibfnamefont {S.}~\bibnamefont
  {Edwards}},\ }\href@noop {} {\bibfield  {journal} {\bibinfo  {journal}
  {Physica A}\ }\textbf {\bibinfo {volume} {353}},\ \bibinfo {pages} {114}
  (\bibinfo {year} {2005})}\BibitemShut {NoStop}%
\bibitem [{\citenamefont {{Edwards}}\ and\ \citenamefont
  {{Oakeshott}}(1989)}]{Edwards89}%
  \BibitemOpen
  \bibfield  {author} {\bibinfo {author} {\bibfnamefont {S.~F.}\ \bibnamefont
  {{Edwards}}}\ and\ \bibinfo {author} {\bibfnamefont {R.~B.~S.}\ \bibnamefont
  {{Oakeshott}}},\ }\href {\doibase 10.1016/0378-4371(89)90034-4} {\bibfield
  {journal} {\bibinfo  {journal} {Physica A}\ }\textbf {\bibinfo {volume}
  {157}},\ \bibinfo {pages} {1080} (\bibinfo {year} {1989})}\BibitemShut
  {NoStop}%
\bibitem [{\citenamefont {Bi}\ \emph {et~al.}(2015)\citenamefont {Bi},
  \citenamefont {Henkes}, \citenamefont {Daniels},\ and\ \citenamefont
  {Chakraborty}}]{Bi15}%
  \BibitemOpen
  \bibfield  {author} {\bibinfo {author} {\bibfnamefont {D.}~\bibnamefont
  {Bi}}, \bibinfo {author} {\bibfnamefont {S.}~\bibnamefont {Henkes}}, \bibinfo
  {author} {\bibfnamefont {K.~E.}\ \bibnamefont {Daniels}}, \ and\ \bibinfo
  {author} {\bibfnamefont {B.}~\bibnamefont {Chakraborty}},\ }\href@noop {}
  {\bibfield  {journal} {\bibinfo  {journal} {Annu. Rev. Condens. Matter
  Phys.}\ }\textbf {\bibinfo {volume} {6}},\ \bibinfo {pages} {63} (\bibinfo
  {year} {2015})}\BibitemShut {NoStop}%
\bibitem [{\citenamefont {Makse}\ and\ \citenamefont
  {Kurchan}(2002)}]{Makse02}%
  \BibitemOpen
  \bibfield  {author} {\bibinfo {author} {\bibfnamefont {H.}~\bibnamefont
  {Makse}}\ and\ \bibinfo {author} {\bibfnamefont {J.}~\bibnamefont
  {Kurchan}},\ }\href@noop {} {\bibfield  {journal} {\bibinfo  {journal}
  {Nature}\ }\textbf {\bibinfo {volume} {415}},\ \bibinfo {pages} {614}
  (\bibinfo {year} {2002})}\BibitemShut {NoStop}%
\bibitem [{\citenamefont {Metzger}(2004)}]{Metzger04}%
  \BibitemOpen
  \bibfield  {author} {\bibinfo {author} {\bibfnamefont {P.~T.}\ \bibnamefont
  {Metzger}},\ }\href@noop {} {\bibfield  {journal} {\bibinfo  {journal}
  {Physical Review E}\ }\textbf {\bibinfo {volume} {70}},\ \bibinfo {pages}
  {051303} (\bibinfo {year} {2004})}\BibitemShut {NoStop}%
\bibitem [{\citenamefont {Asenjo}\ \emph {et~al.}(2014)\citenamefont {Asenjo},
  \citenamefont {Paillusson},\ and\ \citenamefont {Frenkel}}]{Asenjo14}%
  \BibitemOpen
  \bibfield  {author} {\bibinfo {author} {\bibfnamefont {D.}~\bibnamefont
  {Asenjo}}, \bibinfo {author} {\bibfnamefont {F.}~\bibnamefont {Paillusson}},
  \ and\ \bibinfo {author} {\bibfnamefont {D.}~\bibnamefont {Frenkel}},\
  }\href@noop {} {\bibfield  {journal} {\bibinfo  {journal} {Physical review
  letters}\ }\textbf {\bibinfo {volume} {112}},\ \bibinfo {pages} {098002}
  (\bibinfo {year} {2014})}\BibitemShut {NoStop}%
\bibitem [{\citenamefont {Henkes}\ and\ \citenamefont
  {Chakraborty}(2005)}]{Henkes05}%
  \BibitemOpen
  \bibfield  {author} {\bibinfo {author} {\bibfnamefont {S.}~\bibnamefont
  {Henkes}}\ and\ \bibinfo {author} {\bibfnamefont {B.}~\bibnamefont
  {Chakraborty}},\ }\href {\doibase 10.1103/PhysRevLett.95.198002} {\bibfield
  {journal} {\bibinfo  {journal} {Phys. Rev. Lett.}\ }\textbf {\bibinfo
  {volume} {95}},\ \bibinfo {pages} {198002} (\bibinfo {year}
  {2005})}\BibitemShut {NoStop}%
\bibitem [{\citenamefont {Henkes}\ \emph {et~al.}(2007)\citenamefont {Henkes},
  \citenamefont {O'Hern},\ and\ \citenamefont {Chakraborty}}]{Henkes07}%
  \BibitemOpen
  \bibfield  {author} {\bibinfo {author} {\bibfnamefont {S.}~\bibnamefont
  {Henkes}}, \bibinfo {author} {\bibfnamefont {C.~S.}\ \bibnamefont {O'Hern}},
  \ and\ \bibinfo {author} {\bibfnamefont {B.}~\bibnamefont {Chakraborty}},\
  }\href {\doibase 10.1103/PhysRevLett.99.038002} {\bibfield  {journal}
  {\bibinfo  {journal} {Phys. Rev. Lett.}\ }\textbf {\bibinfo {volume} {99}},\
  \bibinfo {pages} {038002} (\bibinfo {year} {2007})}\BibitemShut {NoStop}%
\bibitem [{\citenamefont {Henkes}(2009)}]{Henkes09}%
  \BibitemOpen
  \bibfield  {author} {\bibinfo {author} {\bibfnamefont {S.}~\bibnamefont
  {Henkes}},\ }\emph {\bibinfo {title} {A statistical mechanics framework for
  static granular matter}},\ \href@noop {} {Ph.D. thesis},\ \bibinfo  {school}
  {Brandeis U.} (\bibinfo {year} {2009})\BibitemShut {NoStop}%
\bibitem [{\citenamefont {Blumenfeld}\ and\ \citenamefont
  {Edwards}(2009)}]{Blumenfeld09}%
  \BibitemOpen
  \bibfield  {author} {\bibinfo {author} {\bibfnamefont {R.}~\bibnamefont
  {Blumenfeld}}\ and\ \bibinfo {author} {\bibfnamefont {S.~F.}\ \bibnamefont
  {Edwards}},\ }\href@noop {} {\bibfield  {journal} {\bibinfo  {journal} {The
  Journal of Physical Chemistry B}\ }\textbf {\bibinfo {volume} {113}},\
  \bibinfo {pages} {3981} (\bibinfo {year} {2009})}\BibitemShut {NoStop}%
\bibitem [{\citenamefont {Chakraborty}(2010)}]{Chakraborty10}%
  \BibitemOpen
  \bibfield  {author} {\bibinfo {author} {\bibfnamefont {B.}~\bibnamefont
  {Chakraborty}},\ }\href {\doibase 10.1039/B927435A} {\bibfield  {journal}
  {\bibinfo  {journal} {Soft Matter}\ }\textbf {\bibinfo {volume} {6}},\
  \bibinfo {pages} {2884} (\bibinfo {year} {2010})}\BibitemShut {NoStop}%
\bibitem [{\citenamefont {Blumenfeld}\ \emph {et~al.}(2012)\citenamefont
  {Blumenfeld}, \citenamefont {Jordan},\ and\ \citenamefont
  {Edwards}}]{Blumenfeld12}%
  \BibitemOpen
  \bibfield  {author} {\bibinfo {author} {\bibfnamefont {R.}~\bibnamefont
  {Blumenfeld}}, \bibinfo {author} {\bibfnamefont {J.~F.}\ \bibnamefont
  {Jordan}}, \ and\ \bibinfo {author} {\bibfnamefont {S.~F.}\ \bibnamefont
  {Edwards}},\ }\href@noop {} {\bibfield  {journal} {\bibinfo  {journal}
  {Physical review letters}\ }\textbf {\bibinfo {volume} {109}},\ \bibinfo
  {pages} {238001} (\bibinfo {year} {2012})}\BibitemShut {NoStop}%
\bibitem [{\citenamefont {Wang}\ \emph {et~al.}(2012)\citenamefont {Wang},
  \citenamefont {Song}, \citenamefont {Wang},\ and\ \citenamefont
  {Makse}}]{Wang12}%
  \BibitemOpen
  \bibfield  {author} {\bibinfo {author} {\bibfnamefont {K.}~\bibnamefont
  {Wang}}, \bibinfo {author} {\bibfnamefont {C.}~\bibnamefont {Song}}, \bibinfo
  {author} {\bibfnamefont {P.}~\bibnamefont {Wang}}, \ and\ \bibinfo {author}
  {\bibfnamefont {H.~A.}\ \bibnamefont {Makse}},\ }\href@noop {} {\bibfield
  {journal} {\bibinfo  {journal} {Physical Review E}\ }\textbf {\bibinfo
  {volume} {86}},\ \bibinfo {pages} {011305} (\bibinfo {year}
  {2012})}\BibitemShut {NoStop}%
\bibitem [{\citenamefont {Puckett}\ and\ \citenamefont
  {Daniels}(2013)}]{Puckett13}%
  \BibitemOpen
  \bibfield  {author} {\bibinfo {author} {\bibfnamefont {J.~G.}\ \bibnamefont
  {Puckett}}\ and\ \bibinfo {author} {\bibfnamefont {K.~E.}\ \bibnamefont
  {Daniels}},\ }\href@noop {} {\bibfield  {journal} {\bibinfo  {journal}
  {Physical Review Letters}\ }\textbf {\bibinfo {volume} {110}},\ \bibinfo
  {pages} {058001} (\bibinfo {year} {2013})}\BibitemShut {NoStop}%
\bibitem [{\citenamefont {Bililign}\ \emph {et~al.}(2018)\citenamefont
  {Bililign}, \citenamefont {Kollmer},\ and\ \citenamefont
  {Daniels}}]{Bililign18}%
  \BibitemOpen
  \bibfield  {author} {\bibinfo {author} {\bibfnamefont {E.~S.}\ \bibnamefont
  {Bililign}}, \bibinfo {author} {\bibfnamefont {J.~E.}\ \bibnamefont
  {Kollmer}}, \ and\ \bibinfo {author} {\bibfnamefont {K.~E.}\ \bibnamefont
  {Daniels}},\ }\href@noop {} {\enquote {\bibinfo {title} {Protocol-dependence
  and state variables in the force-moment ensemble},}\ } (\bibinfo {year}
  {2018}),\ \bibinfo {note} {arXiv:1802.09641}\BibitemShut {NoStop}%
\bibitem [{\citenamefont {Barrat}\ \emph {et~al.}(2000)\citenamefont {Barrat},
  \citenamefont {Kurchan}, \citenamefont {Loreto},\ and\ \citenamefont
  {Sellitto}}]{Barrat00}%
  \BibitemOpen
  \bibfield  {author} {\bibinfo {author} {\bibfnamefont {A.}~\bibnamefont
  {Barrat}}, \bibinfo {author} {\bibfnamefont {J.}~\bibnamefont {Kurchan}},
  \bibinfo {author} {\bibfnamefont {V.}~\bibnamefont {Loreto}}, \ and\ \bibinfo
  {author} {\bibfnamefont {M.}~\bibnamefont {Sellitto}},\ }\href@noop {}
  {\bibfield  {journal} {\bibinfo  {journal} {Physical review letters}\
  }\textbf {\bibinfo {volume} {85}},\ \bibinfo {pages} {5034} (\bibinfo {year}
  {2000})}\BibitemShut {NoStop}%
\bibitem [{\citenamefont {Biroli}\ and\ \citenamefont
  {Kurchan}(2001)}]{Biroli01a}%
  \BibitemOpen
  \bibfield  {author} {\bibinfo {author} {\bibfnamefont {G.}~\bibnamefont
  {Biroli}}\ and\ \bibinfo {author} {\bibfnamefont {J.}~\bibnamefont
  {Kurchan}},\ }\href@noop {} {\bibfield  {journal} {\bibinfo  {journal}
  {Physical Review E}\ }\textbf {\bibinfo {volume} {64}},\ \bibinfo {pages}
  {016101} (\bibinfo {year} {2001})}\BibitemShut {NoStop}%
\bibitem [{\citenamefont {Satake}(2004)}]{Satake04}%
  \BibitemOpen
  \bibfield  {author} {\bibinfo {author} {\bibfnamefont {M.}~\bibnamefont
  {Satake}},\ }\href {\doibase DOI: 10.1016/j.ijsolstr.2004.05.046} {\bibfield
  {journal} {\bibinfo  {journal} {Int. J. of Sol. and Struc.}\ }\textbf
  {\bibinfo {volume} {41}},\ \bibinfo {pages} {5775 } (\bibinfo {year}
  {2004})}\BibitemShut {NoStop}%
\bibitem [{\citenamefont {Satake}(1986)}]{Satake86}%
  \BibitemOpen
  \bibfield  {author} {\bibinfo {author} {\bibfnamefont {M.}~\bibnamefont
  {Satake}},\ }in\ \href@noop {} {\emph {\bibinfo {booktitle} {Science on Form:
  Proceedings from the First International Symposium on Science on Form}}},\
  \bibinfo {editor} {edited by\ \bibinfo {editor} {\bibfnamefont
  {G.}~\bibnamefont {Ishizaka}}}\ (\bibinfo  {publisher} {KDK},\ \bibinfo
  {year} {1986})\ pp.\ \bibinfo {pages} {191--199}\BibitemShut {NoStop}%
\bibitem [{\citenamefont {Satake}(1992)}]{Satake92}%
  \BibitemOpen
  \bibfield  {author} {\bibinfo {author} {\bibfnamefont {M.}~\bibnamefont
  {Satake}},\ }\href@noop {} {\bibfield  {journal} {\bibinfo  {journal}
  {International Journal of Engineering Science}\ }\textbf {\bibinfo {volume}
  {30}},\ \bibinfo {pages} {1525} (\bibinfo {year} {1992})}\BibitemShut
  {NoStop}%
\bibitem [{\citenamefont {Satake}(1993)}]{Satake93}%
  \BibitemOpen
  \bibfield  {author} {\bibinfo {author} {\bibfnamefont {M.}~\bibnamefont
  {Satake}},\ }\href {\doibase DOI: 10.1016/0167-6636(93)90028-P} {\bibfield
  {journal} {\bibinfo  {journal} {Mechanics of Materials}\ }\textbf {\bibinfo
  {volume} {16}},\ \bibinfo {pages} {65 } (\bibinfo {year} {1993})},\ \bibinfo
  {note} {special Issue on Mechanics of Granular Materials}\BibitemShut
  {NoStop}%
\bibitem [{\citenamefont {Satake}(1997)}]{Satake97}%
  \BibitemOpen
  \bibfield  {author} {\bibinfo {author} {\bibfnamefont {M.}~\bibnamefont
  {Satake}},\ }in\ \href@noop {} {\emph {\bibinfo {booktitle} {IUTAM Symp. on
  Mech. of Gran. Mat.}}}\ (\bibinfo  {publisher} {Kluwer Academic},\ \bibinfo
  {address} {Dordrecht, The Netherlands},\ \bibinfo {year} {1997})\ pp.\
  \bibinfo {pages} {193 -- 202}\BibitemShut {NoStop}%
\bibitem [{\citenamefont {Ball}\ and\ \citenamefont
  {Blumenfeld}(2002)}]{Ball02}%
  \BibitemOpen
  \bibfield  {author} {\bibinfo {author} {\bibfnamefont {R.~C.}\ \bibnamefont
  {Ball}}\ and\ \bibinfo {author} {\bibfnamefont {R.}~\bibnamefont
  {Blumenfeld}},\ }\href@noop {} {\bibfield  {journal} {\bibinfo  {journal}
  {Phys. Rev. Lett.}\ }\textbf {\bibinfo {volume} {88}},\ \bibinfo {pages}
  {115505} (\bibinfo {year} {2002})}\BibitemShut {NoStop}%
\bibitem [{\citenamefont {DeGiuli}\ and\ \citenamefont
  {McElwaine}(2011)}]{DeGiuli11}%
  \BibitemOpen
  \bibfield  {author} {\bibinfo {author} {\bibfnamefont {E.}~\bibnamefont
  {DeGiuli}}\ and\ \bibinfo {author} {\bibfnamefont {J.}~\bibnamefont
  {McElwaine}},\ }\href {\doibase 10.1103/PhysRevE.84.041310} {\bibfield
  {journal} {\bibinfo  {journal} {Phys. Rev. E}\ }\textbf {\bibinfo {volume}
  {84}},\ \bibinfo {pages} {041310} (\bibinfo {year} {2011})}\BibitemShut
  {NoStop}%
\bibitem [{\citenamefont {DeGiuli}(2013)}]{DeGiuli13}%
  \BibitemOpen
  \bibfield  {author} {\bibinfo {author} {\bibfnamefont {E.}~\bibnamefont
  {DeGiuli}},\ }\emph {\bibinfo {title} {Continuum limits of granular
  systems}},\ \href@noop {} {Ph.D. thesis},\ \bibinfo  {school} {University of
  British Columbia} (\bibinfo {year} {2013})\BibitemShut {NoStop}%
\bibitem [{\citenamefont {DeGiuli}\ and\ \citenamefont
  {Schoof}(2014)}]{DeGiuli14a}%
  \BibitemOpen
  \bibfield  {author} {\bibinfo {author} {\bibfnamefont {E.}~\bibnamefont
  {DeGiuli}}\ and\ \bibinfo {author} {\bibfnamefont {C.}~\bibnamefont
  {Schoof}},\ }\href {http://stacks.iop.org/0295-5075/105/i=2/a=28001}
  {\bibfield  {journal} {\bibinfo  {journal} {EPL (Europhysics Letters)}\
  }\textbf {\bibinfo {volume} {105}},\ \bibinfo {pages} {28001} (\bibinfo
  {year} {2014})}\BibitemShut {NoStop}%
\bibitem [{\citenamefont {Charbonneau}\ \emph {et~al.}(2017)\citenamefont
  {Charbonneau}, \citenamefont {Kurchan}, \citenamefont {Parisi}, \citenamefont
  {Urbani},\ and\ \citenamefont {Zamponi}}]{Charbonneau17}%
  \BibitemOpen
  \bibfield  {author} {\bibinfo {author} {\bibfnamefont {P.}~\bibnamefont
  {Charbonneau}}, \bibinfo {author} {\bibfnamefont {J.}~\bibnamefont
  {Kurchan}}, \bibinfo {author} {\bibfnamefont {G.}~\bibnamefont {Parisi}},
  \bibinfo {author} {\bibfnamefont {P.}~\bibnamefont {Urbani}}, \ and\ \bibinfo
  {author} {\bibfnamefont {F.}~\bibnamefont {Zamponi}},\ }\href@noop {}
  {\bibfield  {journal} {\bibinfo  {journal} {Annual Review of Condensed Matter
  Physics}\ }\textbf {\bibinfo {volume} {8}},\ \bibinfo {pages} {265} (\bibinfo
  {year} {2017})}\BibitemShut {NoStop}%
\bibitem [{\citenamefont {DeGiuli}(2018)}]{atmp_DeGiuli18_short}%
  \BibitemOpen
  \bibfield  {author} {\bibinfo {author} {\bibfnamefont {E.}~\bibnamefont
  {DeGiuli}},\ }\href@noop {} {\enquote {\bibinfo {title} {Field theory for
  amorphous solids},}\ } (\bibinfo {year} {2018}),\ \bibinfo {note} {to appear
  in Phys. Rev. Lett.}\BibitemShut {Stop}%
\bibitem [{\citenamefont {Muskhelishvili}(1963)}]{Muskhelishvili63}%
  \BibitemOpen
  \bibfield  {author} {\bibinfo {author} {\bibfnamefont {N.}~\bibnamefont
  {Muskhelishvili}},\ }\href@noop {} {\emph {\bibinfo {title} {Some basic
  problems of the mathematical theory of elasticity}}}\ (\bibinfo  {publisher}
  {P. Noordhoff},\ \bibinfo {address} {Groningen},\ \bibinfo {year}
  {1963})\BibitemShut {NoStop}%
\bibitem [{\citenamefont {DiDonna}\ and\ \citenamefont
  {Lubensky}(2005)}]{DiDonna05}%
  \BibitemOpen
  \bibfield  {author} {\bibinfo {author} {\bibfnamefont {B.}~\bibnamefont
  {DiDonna}}\ and\ \bibinfo {author} {\bibfnamefont {T.}~\bibnamefont
  {Lubensky}},\ }\href@noop {} {\bibfield  {journal} {\bibinfo  {journal}
  {Physical Review E}\ }\textbf {\bibinfo {volume} {72}},\ \bibinfo {pages}
  {066619} (\bibinfo {year} {2005})}\BibitemShut {NoStop}%
\bibitem [{\citenamefont {Wang}\ and\ \citenamefont
  {Rutqvist}(2013)}]{Wang13a}%
  \BibitemOpen
  \bibfield  {author} {\bibinfo {author} {\bibfnamefont {Y.}~\bibnamefont
  {Wang}}\ and\ \bibinfo {author} {\bibfnamefont {J.}~\bibnamefont
  {Rutqvist}},\ }\href@noop {} {\bibfield  {journal} {\bibinfo  {journal}
  {Journal of Elasticity}\ }\textbf {\bibinfo {volume} {113}},\ \bibinfo
  {pages} {283} (\bibinfo {year} {2013})}\BibitemShut {NoStop}%
\bibitem [{\citenamefont {Gurtin}(1963)}]{Gurtin63}%
  \BibitemOpen
  \bibfield  {author} {\bibinfo {author} {\bibfnamefont {M.~E.}\ \bibnamefont
  {Gurtin}},\ }\href@noop {} {\bibfield  {journal} {\bibinfo  {journal}
  {Archive for Rational Mechanics and Analysis}\ }\textbf {\bibinfo {volume}
  {13}},\ \bibinfo {pages} {321} (\bibinfo {year} {1963})}\BibitemShut
  {NoStop}%
\bibitem [{\citenamefont {Gurtin}(1973)}]{Gurtin73}%
  \BibitemOpen
  \bibfield  {author} {\bibinfo {author} {\bibfnamefont {M.~E.}\ \bibnamefont
  {Gurtin}},\ }\enquote {\bibinfo {title} {The linear theory of elasticity},}\
  in\ \href@noop {} {\emph {\bibinfo {booktitle} {Linear Theories of Elasticity
  and Thermoelasticity}}}\ (\bibinfo  {publisher} {Springer},\ \bibinfo {year}
  {1973})\ pp.\ \bibinfo {pages} {1--295}\BibitemShut {NoStop}%
\bibitem [{\citenamefont {Rostamian}(1979)}]{Rostamian79}%
  \BibitemOpen
  \bibfield  {author} {\bibinfo {author} {\bibfnamefont {R.}~\bibnamefont
  {Rostamian}},\ }\href@noop {} {\bibfield  {journal} {\bibinfo  {journal}
  {Journal of Elasticity}\ }\textbf {\bibinfo {volume} {9}},\ \bibinfo {pages}
  {349} (\bibinfo {year} {1979})}\BibitemShut {NoStop}%
\bibitem [{\citenamefont {Maier}\ \emph {et~al.}(2017)\citenamefont {Maier},
  \citenamefont {Zippelius},\ and\ \citenamefont {Fuchs}}]{Maier17}%
  \BibitemOpen
  \bibfield  {author} {\bibinfo {author} {\bibfnamefont {M.}~\bibnamefont
  {Maier}}, \bibinfo {author} {\bibfnamefont {A.}~\bibnamefont {Zippelius}}, \
  and\ \bibinfo {author} {\bibfnamefont {M.}~\bibnamefont {Fuchs}},\
  }\href@noop {} {\bibfield  {journal} {\bibinfo  {journal} {Physical review
  letters}\ }\textbf {\bibinfo {volume} {119}},\ \bibinfo {pages} {265701}
  (\bibinfo {year} {2017})}\BibitemShut {NoStop}%
\bibitem [{\citenamefont {Bertin}\ \emph {et~al.}(2006)\citenamefont {Bertin},
  \citenamefont {Dauchot},\ and\ \citenamefont {Droz}}]{Bertin06}%
  \BibitemOpen
  \bibfield  {author} {\bibinfo {author} {\bibfnamefont {E.}~\bibnamefont
  {Bertin}}, \bibinfo {author} {\bibfnamefont {O.}~\bibnamefont {Dauchot}}, \
  and\ \bibinfo {author} {\bibfnamefont {M.}~\bibnamefont {Droz}},\ }\href@noop
  {} {\bibfield  {journal} {\bibinfo  {journal} {Physical review letters}\
  }\textbf {\bibinfo {volume} {96}},\ \bibinfo {pages} {120601} (\bibinfo
  {year} {2006})}\BibitemShut {NoStop}%
\bibitem [{\citenamefont {Tighe}\ and\ \citenamefont {Vlugt}(2011)}]{Tighe11a}%
  \BibitemOpen
  \bibfield  {author} {\bibinfo {author} {\bibfnamefont {B.~P.}\ \bibnamefont
  {Tighe}}\ and\ \bibinfo {author} {\bibfnamefont {T.~J.~H.}\ \bibnamefont
  {Vlugt}},\ }\href {http://stacks.iop.org/1742-5468/2011/i=04/a=P04002}
  {\bibfield  {journal} {\bibinfo  {journal} {J. Stat. Mech.}\ }\textbf
  {\bibinfo {volume} {2011}},\ \bibinfo {pages} {P04002} (\bibinfo {year}
  {2011})}\BibitemShut {NoStop}%
\bibitem [{\citenamefont {Wu}\ and\ \citenamefont {Teitel}(2015)}]{Wu15a}%
  \BibitemOpen
  \bibfield  {author} {\bibinfo {author} {\bibfnamefont {Y.}~\bibnamefont
  {Wu}}\ and\ \bibinfo {author} {\bibfnamefont {S.}~\bibnamefont {Teitel}},\
  }\href@noop {} {\bibfield  {journal} {\bibinfo  {journal} {Physical Review
  E}\ }\textbf {\bibinfo {volume} {92}},\ \bibinfo {pages} {022207} (\bibinfo
  {year} {2015})}\BibitemShut {NoStop}%
\bibitem [{\citenamefont {Berthier}\ and\ \citenamefont
  {Biroli}(2011)}]{Berthier11b}%
  \BibitemOpen
  \bibfield  {author} {\bibinfo {author} {\bibfnamefont {L.}~\bibnamefont
  {Berthier}}\ and\ \bibinfo {author} {\bibfnamefont {G.}~\bibnamefont
  {Biroli}},\ }\href@noop {} {\bibfield  {journal} {\bibinfo  {journal}
  {Reviews of Modern Physics}\ }\textbf {\bibinfo {volume} {83}},\ \bibinfo
  {pages} {587} (\bibinfo {year} {2011})}\BibitemShut {NoStop}%
\bibitem [{\citenamefont {Kruyt}\ and\ \citenamefont
  {Rothenburg}(1996)}]{Kruyt96}%
  \BibitemOpen
  \bibfield  {author} {\bibinfo {author} {\bibfnamefont {N.}~\bibnamefont
  {Kruyt}}\ and\ \bibinfo {author} {\bibfnamefont {L.}~\bibnamefont
  {Rothenburg}},\ }\href@noop {} {\bibfield  {journal} {\bibinfo  {journal}
  {Journal of applied mechanics}\ }\textbf {\bibinfo {volume} {63}},\ \bibinfo
  {pages} {706} (\bibinfo {year} {1996})}\BibitemShut {NoStop}%
\bibitem [{\citenamefont {Galley}(2013)}]{Galley13}%
  \BibitemOpen
  \bibfield  {author} {\bibinfo {author} {\bibfnamefont {C.~R.}\ \bibnamefont
  {Galley}},\ }\href@noop {} {\bibfield  {journal} {\bibinfo  {journal}
  {Physical review letters}\ }\textbf {\bibinfo {volume} {110}},\ \bibinfo
  {pages} {174301} (\bibinfo {year} {2013})}\BibitemShut {NoStop}%
\bibitem [{\citenamefont {Tighe}\ \emph {et~al.}(2008)\citenamefont {Tighe},
  \citenamefont {van Eerd},\ and\ \citenamefont {Vlugt}}]{Tighe08}%
  \BibitemOpen
  \bibfield  {author} {\bibinfo {author} {\bibfnamefont {B.~P.}\ \bibnamefont
  {Tighe}}, \bibinfo {author} {\bibfnamefont {A.~R.~T.}\ \bibnamefont {van
  Eerd}}, \ and\ \bibinfo {author} {\bibfnamefont {T.~J.~H.}\ \bibnamefont
  {Vlugt}},\ }\href {\doibase 10.1103/PhysRevLett.100.238001} {\bibfield
  {journal} {\bibinfo  {journal} {Phys. Rev. Lett.}\ }\textbf {\bibinfo
  {volume} {100}},\ \bibinfo {pages} {238001} (\bibinfo {year}
  {2008})}\BibitemShut {NoStop}%
\bibitem [{\citenamefont {Tighe}\ \emph {et~al.}(2010)\citenamefont {Tighe},
  \citenamefont {Snoeijer}, \citenamefont {Vlugt},\ and\ \citenamefont {van
  Hecke}}]{Tighe10a}%
  \BibitemOpen
  \bibfield  {author} {\bibinfo {author} {\bibfnamefont {B.~P.}\ \bibnamefont
  {Tighe}}, \bibinfo {author} {\bibfnamefont {J.~H.}\ \bibnamefont {Snoeijer}},
  \bibinfo {author} {\bibfnamefont {T.~J.~H.}\ \bibnamefont {Vlugt}}, \ and\
  \bibinfo {author} {\bibfnamefont {M.}~\bibnamefont {van Hecke}},\ }\href
  {\doibase 10.1039/B926592A} {\bibfield  {journal} {\bibinfo  {journal} {Soft
  Matter}\ }\textbf {\bibinfo {volume} {6}},\ \bibinfo {pages} {2908} (\bibinfo
  {year} {2010})}\BibitemShut {NoStop}%
\bibitem [{\citenamefont {Tighe}\ and\ \citenamefont {Vlugt}(2010)}]{Tighe10b}%
  \BibitemOpen
  \bibfield  {author} {\bibinfo {author} {\bibfnamefont {B.~P.}\ \bibnamefont
  {Tighe}}\ and\ \bibinfo {author} {\bibfnamefont {T.~J.~H.}\ \bibnamefont
  {Vlugt}},\ }\href {http://stacks.iop.org/1742-5468/2010/i=01/a=P01015}
  {\bibfield  {journal} {\bibinfo  {journal} {J. Stat. Mech.}\ }\textbf
  {\bibinfo {volume} {2010}},\ \bibinfo {pages} {P01015} (\bibinfo {year}
  {2010})}\BibitemShut {NoStop}%
\bibitem [{\citenamefont {Zee}(2010)}]{Zee10}%
  \BibitemOpen
  \bibfield  {author} {\bibinfo {author} {\bibfnamefont {A.}~\bibnamefont
  {Zee}},\ }\href@noop {} {\emph {\bibinfo {title} {Quantum field theory in a
  nutshell}}}\ (\bibinfo  {publisher} {Princeton university press},\ \bibinfo
  {year} {2010})\BibitemShut {NoStop}%
\bibitem [{\citenamefont {Zinn-Justin}(1996)}]{Zinn-Justin96}%
  \BibitemOpen
  \bibfield  {author} {\bibinfo {author} {\bibfnamefont {J.~.~X.}\ \bibnamefont
  {Zinn-Justin}},\ }\href@noop {} {\emph {\bibinfo {title} {Quantum field
  theory and critical phenomena}}}\ (\bibinfo  {publisher} {Clarendon Press},\
  \bibinfo {year} {1996})\BibitemShut {NoStop}%
\bibitem [{\citenamefont {Sadd}(2009)}]{Sadd09}%
  \BibitemOpen
  \bibfield  {author} {\bibinfo {author} {\bibfnamefont {M.~H.}\ \bibnamefont
  {Sadd}},\ }\href@noop {} {\emph {\bibinfo {title} {Elasticity - Theory,
  Applications, and Numerics (2nd Edition)}}}\ (\bibinfo  {publisher}
  {Elsevier},\ \bibinfo {year} {2009})\BibitemShut {NoStop}%
\bibitem [{\citenamefont {Seiberg}(1990)}]{Seiberg90}%
  \BibitemOpen
  \bibfield  {author} {\bibinfo {author} {\bibfnamefont {N.}~\bibnamefont
  {Seiberg}},\ }\href@noop {} {\bibfield  {journal} {\bibinfo  {journal}
  {Progress of Theoretical Physics Supplement}\ }\textbf {\bibinfo {volume}
  {102}},\ \bibinfo {pages} {319} (\bibinfo {year} {1990})}\BibitemShut
  {NoStop}%
\bibitem [{\citenamefont {Andreotti}\ \emph {et~al.}(2013)\citenamefont
  {Andreotti}, \citenamefont {Forterre},\ and\ \citenamefont
  {Pouliquen}}]{Andreotti13}%
  \BibitemOpen
  \bibfield  {author} {\bibinfo {author} {\bibfnamefont {B.}~\bibnamefont
  {Andreotti}}, \bibinfo {author} {\bibfnamefont {Y.}~\bibnamefont {Forterre}},
  \ and\ \bibinfo {author} {\bibfnamefont {O.}~\bibnamefont {Pouliquen}},\
  }\href@noop {} {\emph {\bibinfo {title} {Granular media: between fluid and
  solid}}}\ (\bibinfo  {publisher} {Cambridge University Press},\ \bibinfo
  {year} {2013})\BibitemShut {NoStop}%
\bibitem [{\citenamefont {Strandburg}(1988)}]{Strandburg88}%
  \BibitemOpen
  \bibfield  {author} {\bibinfo {author} {\bibfnamefont {K.~J.}\ \bibnamefont
  {Strandburg}},\ }\href@noop {} {\bibfield  {journal} {\bibinfo  {journal}
  {Reviews of modern physics}\ }\textbf {\bibinfo {volume} {60}},\ \bibinfo
  {pages} {161} (\bibinfo {year} {1988})}\BibitemShut {NoStop}%
\bibitem [{\citenamefont {Beekman}\ \emph {et~al.}(2017)\citenamefont
  {Beekman}, \citenamefont {Nissinen}, \citenamefont {Wu}, \citenamefont {Liu},
  \citenamefont {Slager}, \citenamefont {Nussinov}, \citenamefont {Cvetkovic},\
  and\ \citenamefont {Zaanen}}]{Beekman17}%
  \BibitemOpen
  \bibfield  {author} {\bibinfo {author} {\bibfnamefont {A.~J.}\ \bibnamefont
  {Beekman}}, \bibinfo {author} {\bibfnamefont {J.}~\bibnamefont {Nissinen}},
  \bibinfo {author} {\bibfnamefont {K.}~\bibnamefont {Wu}}, \bibinfo {author}
  {\bibfnamefont {K.}~\bibnamefont {Liu}}, \bibinfo {author} {\bibfnamefont
  {R.-J.}\ \bibnamefont {Slager}}, \bibinfo {author} {\bibfnamefont
  {Z.}~\bibnamefont {Nussinov}}, \bibinfo {author} {\bibfnamefont
  {V.}~\bibnamefont {Cvetkovic}}, \ and\ \bibinfo {author} {\bibfnamefont
  {J.}~\bibnamefont {Zaanen}},\ }\href@noop {} {\bibfield  {journal} {\bibinfo
  {journal} {Physics Reports}\ }\textbf {\bibinfo {volume} {683}},\ \bibinfo
  {pages} {1} (\bibinfo {year} {2017})}\BibitemShut {NoStop}%
\bibitem [{\citenamefont {O'Hern}\ \emph {et~al.}(2003)\citenamefont {O'Hern},
  \citenamefont {Silbert}, \citenamefont {Liu},\ and\ \citenamefont
  {Nagel}}]{OHern03}%
  \BibitemOpen
  \bibfield  {author} {\bibinfo {author} {\bibfnamefont {C.~S.}\ \bibnamefont
  {O'Hern}}, \bibinfo {author} {\bibfnamefont {L.~E.}\ \bibnamefont {Silbert}},
  \bibinfo {author} {\bibfnamefont {A.~J.}\ \bibnamefont {Liu}}, \ and\
  \bibinfo {author} {\bibfnamefont {S.~R.}\ \bibnamefont {Nagel}},\ }\href
  {\doibase 10.1103/PhysRevE.68.011306} {\bibfield  {journal} {\bibinfo
  {journal} {Phys. Rev. E}\ }\textbf {\bibinfo {volume} {68}},\ \bibinfo
  {pages} {011306} (\bibinfo {year} {2003})}\BibitemShut {NoStop}%
\bibitem [{\citenamefont {DeGiuli}\ \emph {et~al.}(2015)\citenamefont
  {DeGiuli}, \citenamefont {D\"uring}, \citenamefont {Lerner},\ and\
  \citenamefont {Wyart}}]{DeGiuli15a}%
  \BibitemOpen
  \bibfield  {author} {\bibinfo {author} {\bibfnamefont {E.}~\bibnamefont
  {DeGiuli}}, \bibinfo {author} {\bibfnamefont {G.}~\bibnamefont {D\"uring}},
  \bibinfo {author} {\bibfnamefont {E.}~\bibnamefont {Lerner}}, \ and\ \bibinfo
  {author} {\bibfnamefont {M.}~\bibnamefont {Wyart}},\ }\href@noop {}
  {\bibfield  {journal} {\bibinfo  {journal} {Physical Review E}\ }\textbf
  {\bibinfo {volume} {91}},\ \bibinfo {pages} {062206} (\bibinfo {year}
  {2015})}\BibitemShut {NoStop}%
\bibitem [{\citenamefont {Gelin}\ \emph {et~al.}(2016)\citenamefont {Gelin},
  \citenamefont {Tanaka},\ and\ \citenamefont {Lema{\^\i}tre}}]{Gelin16}%
  \BibitemOpen
  \bibfield  {author} {\bibinfo {author} {\bibfnamefont {S.}~\bibnamefont
  {Gelin}}, \bibinfo {author} {\bibfnamefont {H.}~\bibnamefont {Tanaka}}, \
  and\ \bibinfo {author} {\bibfnamefont {A.}~\bibnamefont {Lema{\^\i}tre}},\
  }\href@noop {} {\bibfield  {journal} {\bibinfo  {journal} {Nature Materials}\
  } (\bibinfo {year} {2016})}\BibitemShut {NoStop}%
\bibitem [{\citenamefont {Safarov}\ and\ \citenamefont
  {Vassilev}(1997)}]{Safarov97}%
  \BibitemOpen
  \bibfield  {author} {\bibinfo {author} {\bibfnamefont {Y.}~\bibnamefont
  {Safarov}}\ and\ \bibinfo {author} {\bibfnamefont {D.}~\bibnamefont
  {Vassilev}},\ }\href@noop {} {\emph {\bibinfo {title} {The asymptotic
  distribution of eigenvalues of partial differential operators}}},\ Vol.\
  \bibinfo {volume} {155 
  Mathematical Soc.},\ \bibinfo {year} {1997})\BibitemShut {NoStop}%
\bibitem [{\citenamefont {Regge}\ and\ \citenamefont
  {Teitelboim}(1974)}]{Regge74}%
  \BibitemOpen
  \bibfield  {author} {\bibinfo {author} {\bibfnamefont {T.}~\bibnamefont
  {Regge}}\ and\ \bibinfo {author} {\bibfnamefont {C.}~\bibnamefont
  {Teitelboim}},\ }\href@noop {} {\bibfield  {journal} {\bibinfo  {journal}
  {Annals of Physics}\ }\textbf {\bibinfo {volume} {88}},\ \bibinfo {pages}
  {286} (\bibinfo {year} {1974})}\BibitemShut {NoStop}%
\bibitem [{\citenamefont {Hawking}\ and\ \citenamefont
  {Horowitz}(1996)}]{Hawking96}%
  \BibitemOpen
  \bibfield  {author} {\bibinfo {author} {\bibfnamefont {S.~W.}\ \bibnamefont
  {Hawking}}\ and\ \bibinfo {author} {\bibfnamefont {G.~T.}\ \bibnamefont
  {Horowitz}},\ }\href@noop {} {\bibfield  {journal} {\bibinfo  {journal}
  {Classical and Quantum Gravity}\ }\textbf {\bibinfo {volume} {13}},\ \bibinfo
  {pages} {1487} (\bibinfo {year} {1996})}\BibitemShut {NoStop}%
\bibitem [{\citenamefont {Dyer}\ and\ \citenamefont
  {Hinterbichler}(2009)}]{Dyer09}%
  \BibitemOpen
  \bibfield  {author} {\bibinfo {author} {\bibfnamefont {E.}~\bibnamefont
  {Dyer}}\ and\ \bibinfo {author} {\bibfnamefont {K.}~\bibnamefont
  {Hinterbichler}},\ }\href@noop {} {\bibfield  {journal} {\bibinfo  {journal}
  {Physical Review D}\ }\textbf {\bibinfo {volume} {79}},\ \bibinfo {pages}
  {024028} (\bibinfo {year} {2009})}\BibitemShut {NoStop}%
\bibitem [{\citenamefont {Beltrami}(1892)}]{Beltrami92}%
  \BibitemOpen
  \bibfield  {author} {\bibinfo {author} {\bibfnamefont {E.}~\bibnamefont
  {Beltrami}},\ }\href@noop {} {\bibfield  {journal} {\bibinfo  {journal} {Atti
  Accad. Lincei Rend}\ }\textbf {\bibinfo {volume} {1}},\ \bibinfo {pages}
  {141} (\bibinfo {year} {1892})}\BibitemShut {NoStop}%
\end{thebibliography}%

\begin{widetext}

\section{Appendices}
\subsection*{ Appendix 1. Stress tensor correlator and determinant in 3D}
{\blue In 3D, the stress-stress correlator is
\eq{
\langle \sigma_{ij}(\rv) \sigma_{kl}(\rvp) \rangle_c & = \epsilon_{ipq} \epsilon_{jrs} \epsilon_{kuv} \epsilon_{lwx} \p_p \p_r \p'_u \p'_w \langle \Psi_{qs}(\rv) \Psi_{vx}(\rvp) \rangle_c
}
Assuming homogeneity, this can be written
\eq{
\langle \sigma_{ij}(\rv) \sigma_{kl}(0) \rangle_c & = \epsilon_{ipq} \epsilon_{jrs} \epsilon_{kuv} \epsilon_{lwx} \p_p \p_r \p_u \p_w \langle \Psi_{qs}(\rv) \Psi_{vx}(0) \rangle_c
}
In Fourier space this becomes
\eq{
\langle \sigma_{ij}(\vec{q}) \sigma_{kl}(-\vec{q}) \rangle_c & = \epsilon_{ipq} \epsilon_{jrs} \epsilon_{kuv} \epsilon_{lwx} q_p q_r q_u q_w \langle \Psi_{qs}(\vec{q}) \Psi_{vx}(-\vec{q}) \rangle_c
}
From antisymmetry of the Levi-Civita symbol, we see that any contraction along $q_i$, $q_j$, $q_k$, or $q_l$ will vanish. Thus only transverse-transverse stress correlators are nonzero. }

The determinant of $\sigmab$ in 3D is
\eq{
\det \sigma & = \frac{1}{3!} \epsilon_{ijk} \epsilon_{lmn} \sigma_{il} \sigma_{jm} \sigma_{kn} \notag \\ 
& = \frac{1}{3!} \epsilon_{ijk} \epsilon_{lmn} \epsilon_{ipq} \big(\p_p (\nabla \times \Psi)_{lq}\big) \epsilon_{jrs} \big(\p_r (\nabla \times \Psib)_{ms}\big) \epsilon_{ktu} \big(\p_t (\nabla \times \Psi)_{nu}\big)  \notag \\ 
& = \frac{1}{3!} \epsilon_{lmn} \big( \epsilon_{prs} \epsilon_{qtu} - \epsilon_{qrs} \epsilon_{ptu} \big)  \big(\p_p (\nabla \times \Psi)_{lq}\big)  \big(\p_r (\nabla \times \Psib)_{ms}\big) \big(\p_t (\nabla \times \Psi)_{nu}\big)  \notag \\ 
& = \frac{1}{3!} \epsilon_{lmn} \left\{ \big( \epsilon_{prs} \epsilon_{qtu} - \epsilon_{qrs} \epsilon_{ptu} \big) \p_p \left[ (\nabla \times \Psi)_{lq} \big(\p_r (\nabla \times \Psib)_{ms}\big) \big(\p_t (\nabla \times \Psi)_{nu}\big) \right] \right. \\
& \qquad \left. + \epsilon_{prs} \epsilon_{qtu}  (\nabla \times \Psib)_{lq} \p_r (\nabla \times \Psi)_{ms}\p_p \p_t (\nabla \times \Psi)_{nu} \right. \notag \\
& \qquad \left. - \epsilon_{qrs} \epsilon_{ptu} (\nabla \times \Psib)_{lq} \p_p \p_r (\nabla \times \Psi)_{ms}\p_t (\nabla \times \Psi)_{nu}\right\} \notag 
}
{\blue where we integrated by parts. In the final line, two terms have already been eliminated because they involve $\epsilon_{prs} \p_p \p_r$ and $\epsilon_{ptu} \p_p \p_t$, which vanish from antisymmetry of the Levi-Civita symbol. }In the final line, the last two terms are equal and opposite (easily seen after a permutation $r \leftrightarrow t,s \leftrightarrow u$), hence $\det \sigma$ is again a total divergence. It can be simplified to
\eq{
\det \sigma & = \frac{2}{3!} \epsilon_{lmn} \p_p \left[ (\nabla \times \Psib)_{lq} \sigma_{pm} \sigma_{qn} \right]
}

\subsection*{Appendix 2. Partition function in 2D}
We want to compute
\eq{
Z = \int \Dpsi \;e^{-S}, \qquad\;\; S = \int_{\Omega} dV \; \Lc[\psi], 
}
with
\eq{
\Lc[\psi] & = \alphab : \sigmab + \tilde\gamma \det \sigmab + \half \tilde\eta \; \tr^2 \sigmab
}
{\blue
To simplify expressions involving $\sigmab$, notice that $\sigmab=\nabla \times \nabla \times \psi = (\epsb \cdot \nabla) (\epsb \cdot \nabla) \psi = \epsb \cdot (\nabla \nabla \psi) \cdot \epsb^t,$ so that det$\sigmab = \det \epsb \det \nabla \nabla \psi \det \epsb^t = \det \nabla \nabla \psi$, using $\epsb^t = \epsb^{-1}$, and tr $\sigmab = $tr $ \nabla \nabla \psi = \nabla^2 \psi$, using the cyclic property of the trace. Also we have $\alphab : \sigmab = \alphab : (\epsb^t \cdot \nabla \nabla \psi \cdot \epsb) = (\epsb \cdot \alphab \cdot \epsb^t) : (\nabla \nabla \psi) = \alphac : (\nabla \nabla \psi)$ with our definition of $\check{A} = \epsb \cdot \hat{A} \cdot \epsb^t$ for any 2x2 matrix $A$. Thus we find} 
\eq{
\Lc[\psi] = \alphac : (\nabla \nabla \psi) + \tilde\gamma \det \nabla \nabla \psi + \half \tilde\eta \; (\nabla^2 \psi)^2.
}
Although in the physical case $\alphab$ is constant, we can let $\alphab = \hat{\overline{\alpha}} + \alphab_g(\rv)$ where the former controls the mean stress and the latter generates correlation functions. We can write
\eq{
\Lc[\psi] & = \nabla \cdot \big[ \alphac \cdot \nabla \psi + \half \tilde\gamma \sigmab \cdot \nabla \psi + \half \tilde\eta \;  (\nabla^2 \psi) \cdot \nabla \psi \big] - (\nabla \cdot \alphac ) \cdot \nabla \psi - \half \tilde\eta \; \nabla (\nabla^2 \psi) \cdot \nabla \psi \notag \\
& = \nabla \cdot \big[ \alphac \cdot \nabla \psi + \half \tilde\gamma \sigmab \cdot \nabla \psi + \half \tilde\eta \;  (\nabla^2 \psi) \cdot \nabla \psi - (\nabla \cdot \alphac ) \psi - \half \tilde\eta \; \nabla (\nabla^2 \psi) \big] + \psi \big[ \nabla \nabla : \alphac + \half \tilde\eta \nabla^4 \psi \big] \notag \\
& = \nabla \cdot \Jv[\psi,\alphab] + \psi \big[ \nabla \nabla : \alphac + \half \tilde\eta \nabla^4 \psi \big]
}
with
\eq{
\Jv[\psi,\alphab] = \big( \alphac + \half \tilde\gamma \sigmab + \half \tilde\eta \;  \delb (\nabla^2 \psi) \big) \cdot \nabla \psi - \big( \nabla \cdot \alphac + \half \tilde\eta \nabla (\nabla^2 \psi) \big) \psi
}
Now we let $\psi = \psi_c + \psi'$ and find
\eq{
\Lc[\psi] = \nabla \cdot \Jv_c + \nabla \cdot \Jv' + \nabla \cdot \Delta\Jv + \psi_c \big[ \nabla \nabla : \alphac + \half \tilde\eta \nabla^4 \psi_c \big] + \half \tilde\eta \psi' \nabla^4 \psi' + \psi' \big[ \nabla \nabla : \alphac + \tilde\eta \nabla^4 \psi_c \big]   
}
where $\Jv_c = \Jv[\psi_c,\alphab], \Jv' = \Jv[\psi',0]$ and 
\eq{
\Delta \Jv = \big( \alphac + \tilde\gamma \sigmab_c + \tilde\eta \;  (\nabla^2 \psi_c) \big) \cdot \nabla \psi' - \big( \nabla \cdot \alphac + \tilde\eta \nabla (\nabla^2 \psi_c) \big) \psi'
}
To eliminate coupling between $\psi_c$ and $\psi'$ we would like to choose the non-fluctuating `classical' part $\psi_c$ to satisfy
\eq{ \label{app_eom1}
\nabla^4 \psi_c = -\tilde\eta^{-1} \nabla \nabla : \alphac
}
with boundary conditions
\eq{ \label{app_bc1}
0 & = \nv \cdot \left[ \alphac +  \tilde\gamma \sigmab_c +  \tilde\eta \; \delb (\nabla^2 \psi_c) \right], \\
0 & = \nv \cdot \left[ \nabla \cdot \alphac + \tilde\eta \nabla (\nabla^2 \psi_c) \right] \label{app_bc2}
}
where $\nv$ is a boundary normal.  We write $\psi_c = \psibar + \psi_{g}$ where $\sigmabar = \nabla \times \nabla \times \psibar$ is a constant. We see that in order to cancel the term $\check{\overline{\alpha}}$ in the boundary conditions, $\sigmabar$ must satisfy $0 = \check{\overline{\alpha}} +  \tilde\gamma \sigmabar +  \tilde\eta \delb \;\tr\; \sigmabar$, which leads to the equation of state shown in the main text. The second boundary condition \eqref{app_bc2} is identically satisfied for $\psibar$. 

The correlation function is determined by the particular solution $\psi_g$. Existence of such a $\psi_g$ is not guaranteed, because the biharmonic equation generally has solutions only when two DOF are specified on the boundary \cite{Muskhelishvili63}. We return to this point below.

Assuming existence of $\psi_c$ satisfying \eqref{app_eom1}, \eqref{app_bc1}, \eqref{app_bc2}, then having eliminated cross-coupling between $\psi_c$ and $\psi'$, we have simply
\eq{
\Lc[\psi] & = \nabla \cdot \Jv_c + \nabla \cdot \Jv' + \half \psi_c \nabla \nabla : \alphac + \half \tilde\eta \psi' \nabla^4 \psi' \notag \\
& = \half \sigmab_c : \alphab + \nabla \cdot \Jv' + \half \tilde\eta \psi' \nabla^4 \psi'.
}
Then since the change of variable $\psi \to \psi'$ has unit jacobian, the partition function can be written
\eq{
Z = e^{-S_c} \int \Dpsi e^{-S'}, \;\; S' = \int_\Omega dV \; \left[ \nabla \cdot \Jv' + \half \tilde\eta \psi' \nabla^4 \psi' \right]
}
where
\eq{
S_c & =  \half \int dV \; \left[ \nabla \cdot [ \alphac \cdot \nabla \psi_c - \psi_c \nabla \cdot \alphac ] + \psi_c \nabla \nabla : \alphac \right] \notag \\
& = \half \int dV \; \alphac : \nabla \nabla \psi_c \\
& = \half \int dV \; \alphab : \sigmab_c
}
The fluctuating part gives a functional determinant
\eq{
\int \Dpsi \;e^{-S'} = \det{}^{-1/2} (\tilde\eta \nabla^4) \equiv e^{-\frac{1}{2} \mbox{tr} \log \tilde\eta \nabla^4},
}
where in principle boundary conditions should be applied such that $\Jv'$ vanishes on the boundary. In fact the boundary contribution to the functional determinant is sub-extensive \cite{Safarov97} and can be neglected. Then Tr$\; \log \tilde\eta \nabla^4$ is easily evaluated in a Fourier basis \cite{Zee10}:
\eq{
\mbox{Tr} \log \tilde\eta \nabla^4 \equiv V \int \frac{d^d q}{(2\pi)^d} \log (\tilde\eta q^4) = \frac{V\Lambda^2}{4\pi} \log \frac{\tilde\eta \Lambda^4}{e^2},
}
{\blue where $\Lambda$ is a UV cutoff in Fourier space, }and the last equation holds only in $d=2$. The final result is
\eq{
\log Z = -S_c - \frac{V\Lambda^2}{8\pi}  \log \frac{\tilde\eta \Lambda^4}{e^2} + \OO(\sqrt{V}),
}
Finally, let us mention how to resolve the apparent mismatch in the number of boundary conditions needed to apply to $\psi_g$. Suppose that 
\eq{ \label{app_h1}
\nv \cdot \left[ \nabla \cdot \alphac + \tilde\eta \nabla (\nabla^2 \psi_g) \right] = \nv \cdot \epsb \cdot \nabla h,
}
for some function $h(\rv)$. Then we can write
\eq{
\int dV \; \nabla \cdot \left[ \left( \nabla \cdot \alphac + \tilde\eta \nabla (\nabla^2 \psi_g) \right) \psi' \right] & = \int ds \;\nv \cdot \left[ \nabla \cdot \alphac + \tilde\eta \nabla (\nabla^2 \psi_g) \right] \psi' \notag \\
& = \int ds \;\nv \cdot \epsb \cdot (\nabla h) \psi' \notag \\
& = \int ds \;\nv \cdot \epsb \cdot \nabla (h \psi') - \int ds \nv \cdot \epsb \cdot h \nabla \psi' \notag \\
& = 0 - \int ds \;\nv \cdot \left[ \epsb \cdot h \nabla \psi' \right],
}
where we used the fact that $\nv \cdot \epsb$ is a vector along the boundary, so that the gradient theorem implies $\int ds \;\nv \cdot \epsb \cdot \nabla (h \psi')=0$. We see that the final term adds to the boundary condition conjugate to $\nabla \psi'$, i.e. \eqref{app_bc1} becomes
\eq{ \label{app_bc3}
0 = \nv \cdot \left[ \alphac +  \tilde\gamma \sigmab_c +  \tilde\eta \; \delb (\nabla^2 \psi_c) + h \; \epsb \right]
}
The equation \eqref{app_h1} can be used to determine the function $h$ by integration (provided the left hand side of \eqref{app_h1} integrates to zero around the boundary). Then \eqref{app_bc2} is no longer relevant, and we have simply \eqref{app_bc3}, which is the correct number of boundary conditions for the biharmonic equation. 

If \eqref{app_bc3} leads to an ill-posed boundary value problem, then it is possible that boundary terms need to be added to the action so that spurious boundary conditions are eliminated \cite{Regge74}. This situation occurs in the treatment of general relativity in finite domains, for which the Einstein-Hilbert action needs to be supplemented by the Gibbons-Hawking-York boundary term to make the action principle well-posed \cite{Hawking96,Dyer09}.

\subsection*{Appendix 3. Partition function in 3D. } We want to compute
\eq{
Z = \int \DPsi \;e^{-S}, \qquad\;\; S = \int_{\Omega} dV \; \Lc[\Psi], 
}
Although $\Psi$ has a nontrivial gauge freedom, this gauge group is abelian, so there is no need to introduce Faddeev-Popov ghosts \cite{Zee10}; we can simply fix the Maxwell gauge. We have
\eq{
\Lc[\Psi] & = \alphab : \sigmab + \gamma \det \sigmab + \half \eta \; \tr^2 \sigmab + \half g \; \tr\;\sigmab\cdot \sigmab \\
& = \alpha_{ij} : (\epsilon_{ikl} \epsilon_{jmn} \p_k \p_m \Psi_{ln}) + \ffrac{1}{3} \gamma \;\epsilon_{lmn} \p_p \left[ (\nabla \times \Psi)_{lq} \sigma_{pm} \sigma_{qn} \right] + \half \eta \; (\epsilon_{ikl} \epsilon_{imn} \p_k \p_m \Psi_{ln} \sigma_{jj}) + \half g \; (\epsilon_{ikl} \epsilon_{jmn} \p_k \p_m \Psi_{ln} \sigma_{ij}) \notag \\
& = \p_k \left( \alpha_{ij} (\epsilon_{ikl} \epsilon_{jmn} \p_m \Psi_{ln}) \right) - (\p_k \alpha_{ij}) : (\epsilon_{ikl}  \epsilon_{jmn} \p_m \Psi_{ln}) +  \ffrac{1}{3} \gamma \;\epsilon_{lmn} \p_p \left[ (\nabla \times \Psi)_{lq} \sigma_{pm} \sigma_{qn} \right] \notag \\
& \qquad + \half \eta \; \p_k (\epsilon_{ikl} \epsilon_{imn} \p_m \Psi_{ln} \sigma_{jj}) - \half \eta \; (\epsilon_{ikl} \epsilon_{imn} \p_m \Psi_{ln} \p_k \sigma_{jj})  \notag \\
& \qquad + \half g \; \p_k (\epsilon_{ikl} \epsilon_{jmn} \p_m \Psi_{ln} \sigma_{ij}) - \half g \; (\epsilon_{ikl} \epsilon_{jmn} \p_m \Psi_{ln} \p_k \sigma_{ij})
}
Continuing, we have
\eq{
(\p_k \alpha_{ij}) (\epsilon_{ikl}  \epsilon_{jmn} \p_m \Psi_{ln}) = \p_m \left( (\p_k \alpha_{ij}) (\epsilon_{ikl}  \epsilon_{jmn} \Psi_{ln}) \right) - (\p_m \p_k \alpha_{ij}) (\epsilon_{ikl}  \epsilon_{jmn}  \Psi_{ln}) 
}
and
\eq{
& \epsilon_{ikl} \epsilon_{imn} \p_m \Psi_{ln} \p_k \sigma_{jj} = \p_m (\epsilon_{ikl} \epsilon_{imn} \Psi_{ln} \p_k \sigma_{jj}) - \epsilon_{ikl} \epsilon_{imn} \Psi_{ln} \p_m \p_k \sigma_{jj} \notag \\
& \epsilon_{ikl} \epsilon_{jmn} \p_m \Psi_{ln} \p_k \sigma_{ij} = \p_m (\epsilon_{ikl} \epsilon_{jmn} \Psi_{ln} \p_k \sigma_{ij}) - \epsilon_{ikl} \epsilon_{jmn} \Psi_{ln} \p_m \p_k \sigma_{ij}
}
Thus we can write
\eq{
\Lc[\Psi] = \nabla \cdot \Jv[\Psi,\alpha] + \Psi : A[\Psi,\alpha] 
}
with
\eq{
J_k[\Psi,\alpha] & = \alpha_{ij} (\epsilon_{ikl} \epsilon_{jmn} \p_m \Psi_{ln}) - (\p_m \alpha_{ij}) (\epsilon_{iml}  \epsilon_{jkn} \Psi_{ln}) +  \ffrac{1}{3} \gamma \;\epsilon_{lmn} (\nabla \times \Psi)_{lq} \sigma_{km} \sigma_{qn} \notag \\
& \qquad + \half \eta \epsilon_{ikl} \epsilon_{imn} \p_m \Psi_{ln} \sigma_{jj} - \half \eta \epsilon_{iml} \epsilon_{ikn} \Psi_{ln} \p_m \sigma_{jj} \notag \\
& \qquad + \half g \epsilon_{ikl} \epsilon_{jmn} \p_m \Psi_{ln} \sigma_{ij} - \half g \epsilon_{iml} \epsilon_{jkn} \Psi_{ln} \p_m \sigma_{ij}
}
and
\eq{
A_{ij}[\Psi,\alpha] = \epsilon_{lmi} \p_m \p_k \left[ \alpha_{ln} \epsilon_{nkj}   + \half \eta \;\epsilon_{lkj}  \sigma_{nn} + \half g \;\epsilon_{nkj}  \sigma_{ln} \right]
}
After some work this can be written
\eq{
A_{ij}[\Psi,\alpha] = \epsilon_{lmi} \epsilon_{nkj} \p_m \p_k \alpha_{ln} + \half (\eta+g) \delta_{ij} \nabla^2 \sigma_{kk} - \half (\eta+g) \p_i \p_j \sigma_{kk} - \half g \nabla^2 \sigma_{ij}
}
Now we perform the same steps as in the 2D case: we let $\alphab = \hat{\overline{\alpha}} + \alphab_g(\rv)$ where the former controls the mean stress and the latter generates correlation functions, and we write $\Psi = \Psi_c + \Psi'$. Then
\eq{
\Lc[\Psi] = \nabla \cdot \Jv[\Psi_c,\alpha] + \nabla \cdot \Jv[\Psi',0]  + \nabla \cdot \Delta\Jv + \Psi_c : A[\Psi_c,\alpha] + \Psi' : A[\Psi',0]  + \Psi' : A[2\Psi_c,\alpha] 
}
with
\eq{
\Delta \Jv_k & = \alpha_{ij} (\epsilon_{ikl} \epsilon_{jmn} \p_m \Psi'_{ln}) - (\p_m \alpha_{ij}) (\epsilon_{iml}  \epsilon_{jkn} \Psi'_{ln}) + \gamma \epsilon_{lmn} (\nabla \times \Psi')_{lq} \sigma_{c,km} \sigma'_{qn} + \gamma \epsilon_{lmn} (\nabla \times \Psi')_{lq} \sigma_{c,km} \sigma_{c,qn} \notag \\
& \qquad +  \eta \;\epsilon_{ikl} \epsilon_{imn} \p_m \Psi'_{ln} \sigma_{c,jj} -  \eta \;\epsilon_{iml} \epsilon_{ikn} \Psi'_{ln} \p_m \sigma_{c,jj} \notag \\
& \qquad +  g \;\epsilon_{ikl} \epsilon_{jmn} \p_m \Psi'_{ln} \sigma_{c,ij} -  g \;\epsilon_{iml} \epsilon_{jkn} \Psi'_{ln} \p_m \sigma_{c,ij}
}
Since power counting indicates that only terms quadratic in the fields are needed at large scales, we can take the semi-classical limit in which $(\nabla \times \Psi')_{lq} \sigma_{c,km} \sigma'_{qn}$ is sub-dominant, being quadratic in the fluctuations. Then the classical solution is given by solving $A[2\Psi_c,\alpha]=0$ subject to boundary conditions
\eq{
0 &= n_k \left[ \alpha_{ij} \epsilon_{ikl}  +  \gamma \; \epsilon_{jmn} \sigma_{c,km} \sigma_{c,ln} +  \eta \; \epsilon_{jkl} \sigma_{c,ii} +  g \; \epsilon_{ikl} \sigma_{c,ij} \right] \\
0 & = n_k \epsilon_{iml} \p_m \left[  \alpha_{ij} \epsilon_{jkn}  + \eta \;\epsilon_{ikn}  \sigma_{c,jj} + g \;\epsilon_{jkn}  \sigma_{c,ij} \right], 
}
where $n_k$ is a boundary normal. We write $\Psi_c = \overline{\Psi} + \Psi_g$, where $\sigmabar = \nabla \times \nabla \times \overline{\Psi}$ is constant. It is fixed by the equation of state
\eq{
0 = \alpha_{ij} + \eta \; \delta_{ij} \sigmabar_{kk} + g \; \sigmabar_{ij} + \half \gamma \; \epsilon_{ikl} \epsilon_{jmn} \sigmabar_{km} \sigmabar_{ln}
}
After some manipulations $Z$ can be written
\eq{
Z = e^{-S_c} \int \DPsi \;e^{-S'}
}
with
\eq{
S_c = \half \int dV \; \left[ \alphab: \sigmab_c - \gamma |\sigmab| \right]
}

To find the stress correlator, we write $A[2\Psi_g,\alpha_g]=0$ as 
\eq{
0 = \nabla \times \nabla \times [ \alphab_g + \eta \delb \;\tr \;\sigmab + g \;\sigmab]
}
which is solved in simply connected domains by
\eq{ \label{stvenant}
\alphab_g + \eta \delb \;\tr \;\sigmab_g + g \;\sigmab_g = \nabla \uv + (\nabla \uv)^t,
}
as shown by Beltrami \cite{Beltrami92}. Then
\eq{
2 \nabla \cdot \uv = \tr \;\alphab_g + (3\eta+g) \tr \sigmab_g
}
and 
\eq{
\nabla^2 \uv + \nabla \nabla \cdot \uv = \nabla \cdot \alphab_g + \frac{\eta}{3\eta+g} \nabla \left[ 2 \nabla \cdot \uv - \tr \;\alphab_g \right]
}
Taking a divergence,
\eq{
2 \frac{2\eta+g}{3\eta+g} \nabla^2 \nabla \cdot \uv = \nabla \nabla : \alphab_g - \frac{\eta}{3\eta+g} \nabla^2 \tr \;\alphab_g
}
For a source $\alphab_g = \alphab_0 \delta(\rv) = -\frac{1}{4\pi} \alphab_0 \nabla^2 (1/r)$ we see that
\eq{
2 \frac{2\eta+g}{3\eta+g} \nabla \cdot \uv = \frac{-1}{4\pi} \left[ \alphab_0 : \nabla \nabla - \frac{\eta \tr\;\alphab_0}{3\eta+g} \nabla^2 \right] \frac{1}{r} + h_1(\rv),
}
where $\nabla^2 h_1 = 0$. Then we have
\eq{
\uv & = \vec{h}_2 + \frac{1}{\nabla^2} \left[ \nabla \cdot \alphab_g -  \frac{\eta}{3\eta+g} \nabla \tr \alphab_g -  \frac{\eta+g}{3\eta+g} \nabla \nabla \cdot \uv \right], \notag \\
& = \vec{h}_2 - \frac{1}{4\pi} \left[ \alphab_0 - \frac{\eta}{3\eta+g} \delb \; \tr \;\alphab_0 \right] \cdot \nabla \frac{1}{r}  - \frac{\eta+g}{3\eta+g} \frac{1}{\nabla^2} \nabla \nabla \cdot \uv,
}
where $\nabla^2 \vec{h}_2(\rv)=0$. Using \eqref{stvenant} we obtain $\sigmab_g$. The harmonic functions $h_1$ and $\vec{h}_2$ can be used to satisfy boundary conditions.

\subsection*{Appendix 4. Spatial versus ensemble fluctuations}
Define the ensemble and spatial fluctuations as
\eq{
C_e = \left\langle \left(\overline{p - \langle p \rangle} \right)^2 \right\rangle, \qquad C_s = \left\langle \overline{ (p - \overline{p})^2 } \right\rangle,
}
Note that the full pressure fluctuations are their sum:
\eq{
\left\langle \overline{ \left(p - \langle \overline{p} \rangle \right)^2 } \right\rangle = C_e + C_s
}
One easily sees that
\eq{
\frac{\p \log Z}{\p \alphab} = - V \langle \overline{\sigmab} \rangle, \qquad \frac{\p^2 \log Z}{\p \alphab \p \alphab} = V^2 \langle \overline{\sigmab \sigmab} \rangle_c, \qquad \frac{\p \log Z}{\p \eta} = - \half V \left\langle \overline{\tr^2 \sigmab} \right\rangle.
}
Let us write $\alphab = \alpha \delb + \fsl{\alphab}$ where $\tr \; \fsl{\alphab} = 0$. Then
\eq{ 
C_e & = \frac{1}{d^2 V^2} \frac{\p^2 \log Z}{\p \alpha^2} \label{fluc1} \\
C_s & = \frac{-2}{d^2 V} \frac{\p \log Z}{\p \eta} - C_e - \frac{1}{d^2 V^2} \left( \frac{\p \log Z}{\p \alpha} \right)^2 \label{fluc2}
}
For $d=2$, we have
\eq{
C_e = \frac{1}{2 V (2\eta+g+\gamma)}, \qquad C_s = \frac{\Lambda^2}{16 \pi (\eta+g)} - C_e
}
In $d=3$, evaluating $C_s$ would require computation of the functional determinant. 

\subsection*{Appendix 5. Strictly repulsive forces}

We have to add terms $-\nu \log p[\psi]$ and $\half m |\nabla \log p[\psi]|^2$ to the action. In both 2D and 3D we have
 \eq{
-\log p[\psi_c + \psi'] & =  -\log p[\psi_c] - \log \left( 1 + \frac{p[\psi']}{p[\psi_c]} \right) \notag \\
 & =  -\log p[\psi_c] -  \frac{p[\psi']}{p[\psi_c]} + \half  \frac{p[\psi']^2}{p[\psi_c]^2} + \ldots,
}
where we have assumed $|p[\psi']|<p[\psi_c]$, necessary for physical solutions. In 2D we have $p=\half \nabla^2 \psi$ so that
\eq{
 - \log p[\psi_c + \psi'] +  \log p_c & = \nabla \cdot \left[ -\frac{1}{2p_c} \nabla \psi' + \frac{1}{8 p_c^2} \nabla^2 \psi' \nabla \psi' \right] + \nabla \left(\frac{1}{2p_c}\right) \cdot \nabla \psi' - \nabla \left(\frac{1}{8 p_c^2} \nabla^2 \psi' \right) \cdot \nabla \psi' + \ldots \notag \\
& = \nabla \cdot \left[ -\frac{1}{2p_c} \nabla \psi' + \frac{1}{8 p_c^2} \nabla^2 \psi' \nabla \psi' 
+ \nabla \left(\frac{1}{2p_c}\right) \psi' - \nabla \left(\frac{1}{8 p_c^2} \nabla^2 \psi' \right) \psi' \right] \notag \\
& \qquad - \nabla^2 \left(\frac{1}{2p_c}\right) \psi' + \nabla^2 \left(\frac{1}{8 p_c^2} \nabla^2 \psi' \right) \psi' + \ldots 
}
Similarly, in both 2D and 3D:
\eq{
|\nabla \log(p_c+p') |^2  & = \left| \nabla \log p_c + \nabla \log \left(1 + \frac{p'}{p_c} \right) \right|^2 \notag \\
& = |\nabla \log p_c|^2 + 2 (\nabla \log p_c) \cdot \nabla \frac{p'}{p_c} + \nabla \left(\frac{p'}{p_c} \right) \cdot \nabla \left( \frac{p'}{p_c} \right) - (\nabla \log p_c) \cdot \nabla \left(\frac{p'{}^2}{p_c^2} \right) + \ldots
}
Writing $L_c = \log p_c$, we have, in 2D,
\eq{
(\nabla \log p_c) \cdot \nabla \frac{p'}{p_c} & = \nabla \cdot \left[ \nabla L_c  \frac{p'}{p_c} - \frac{\nabla^2 L_c}{2 p_c} \nabla \psi' + \nabla \left( \frac{\nabla^2 L_c}{2 p_c} \right) \psi' \right] - \psi' \nabla^2 \left( \frac{\nabla^2 L_c}{2 p_c} \right) \\
\nabla \left( \frac{p'}{p_c} \right) \cdot \nabla \left( \frac{p'}{p_c} \right) & = \nabla \cdot \left[ \frac{p'}{p_c} \cdot \nabla \left( \frac{p'}{p_c} \right) - \frac{\nabla \psi'}{2 p_c} \nabla^2 \left( \frac{p'}{p_c} \right) + \psi' \nabla \left( \frac{1}{2 p_c} \nabla^2 \left( \frac{p'}{p_c} \right) \right) \right] - \psi' \nabla^2 \left( \frac{1}{2 p_c} \nabla^2 \left( \frac{p'}{p_c} \right) \right) \\
\nabla L_c \cdot \nabla \left( \frac{p'{}^2}{p_c^2} \right) & = \nabla \cdot \left[ \frac{p'{}^2}{p_c^2} \nabla L_c - \frac{p' \nabla \psi' }{2 p_c^2} \nabla^2 L_c + \psi'  \nabla \left( \frac{p'}{2 p_c^2} \nabla^2 L_c \right) \right] - \psi' \nabla^2 \left( \frac{p'}{2 p_c^2} \nabla^2 L_c \right) 
}
Since the theory is now nonlinear, cross-coupling between $\psi_c$ and $\psi'$ cannot be entirely eliminated. The best we can do is to eliminate coupling at first order in the fluctuations $\psi'$, so that the field equation becomes
\eq{ \label{app_class}
0 = \nabla \nabla : \alphac + \tilde\eta \nabla^4 \psi_c - \nu \nabla^2 \left(\frac{1}{\nabla^2 \psi_c}\right) - m \nabla^2 \left( \frac{\nabla^2 \log p_c}{2 p_c} \right)
}
with boundary conditions
\eq{ \label{app_bc4}
0 & = \nv \cdot \left[ \alphac +  \tilde\gamma \sigmab_c +  2 \tilde\eta \; \delb \; p_c - \delb \frac{\nu}{2p_c} -m \delb \frac{\nabla^2 \log p_c}{2p_c} \right], \\
0 & = \nv \cdot \left[ \nabla \cdot \alphac + 2 \tilde\eta \nabla p_c - \nu \nabla \left(\frac{1}{2p_c}\right) - m \nabla \left( \frac{\nabla^2 \log p_c}{2p_c}\right) \right] \label{app_bc5} \\
0 & = \nv \cdot \left[ \nabla \log p_c \right] \label{app_bc6}
}
From this we read off the equation of state
\eq{
0 = \alphac +  \tilde\gamma \sigmabbar +  2 \tilde\eta \; \delb \; \overline{p}  - \delb \; \frac{\nu}{2 \overline{p}}
}
The partition function becomes
\eq{
Z = e^{-S_c} \int \Dpsi \; e^{-S'}
}
with
\eq{
S' & = \int dV \; \left[ \nabla \cdot \tilde\Jv' + \half \tilde\eta \psi' \nabla^4 \psi' + \psi' \nabla^2 \left(\frac{\nu}{8 p_c^2} \nabla^2 \psi' \right) - \half m \psi' \nabla^2 \left( \frac{1}{2 p_c} \nabla^2 \left( \frac{p'}{p_c} \right) \right) + \half m \psi' \nabla^2 \left( \frac{p'}{2 p_c^2} \nabla^2 \log p_c \right) \right] \\
S_c & = \int dV \; \left[ \alphab : \sigmab_c + \tilde\gamma |\sigmab_c| + \half \tilde\eta \tr^2 \sigmab - \nu \log p_c + \half m |\nabla \log p_c|^2 \right] 
}
where $\tilde \Jv'$ collects all the boundary fluxes quadratic in $\psi'$. 
The fluctuations give
\eq{
\int \Dpsi \;e^{-S'} & = \det{}^{-1/2} \left[\tilde\eta \nabla^4 + \nu \nabla^2 ((2p_c)^{-2} \nabla^2) - m \nabla^2((2p_c)^{-1} \nabla^2 (2p_c)^{-1} ) + m \nabla^2 (2p_c)^{-2} (\nabla^2 \log p_c)\right] \notag \\
& \equiv e^{-\frac{1}{2} \mbox{tr} \log [\tilde\eta \nabla^4 + \nu \nabla^2 ((2p_c)^{-2} \nabla^2)) - m \nabla^2((2p_c)^{-1} \nabla^2 (2p_c)^{-1} ) + m \nabla^2 (2p_c)^{-2} (\nabla^2 \log p_c)]},
}
where, for example, $\nabla^2 ((2p_c)^{-2} \nabla^2)$ acting on $f$ is  $\nabla^2 \left[ (2p_c)^{-2} \nabla^2 f \right]$. In principle boundary conditions should be applied such that the fluxes vanish, but these are expected to be sub-extensive in large systems. The determinant is nontrivial when $p_c$ is non-constant, in particular when computing the correlator. At leading order we replace $p_c$ in $S'$ by $\overline{p}$, giving
\eq{
\mbox{Tr} \log \left[ \tilde\eta_R \nabla^4 (1 - \xi^2 \nabla^2 ) \right],
}
where $\tilde\eta_R = \tilde\eta + \nu/(4\pbar^2)$ is the renormalized $\tilde\eta$ and $\xi = \sqrt{\frac{m}{4 \tilde\eta_R \pbar^2}}$ defines a length scale. Defining the functional determinant in the Fourier basis, we find
\eq{
\mbox{Tr} \log \left[ \tilde\eta_R \nabla^4 - \frac{m}{4 \pbar^2} \nabla^6 \right] & \equiv V \int \frac{d^d q}{(2\pi)^d} \log \left[ \tilde\eta_R q^4 + \frac{m}{4 \pbar^2} q^6 \right] \notag \\
& = \frac{V\Lambda^2}{4\pi} \log \frac{\tilde\eta_R \Lambda^4}{e^3} + \frac{V (1 + \xi^2 \Lambda^2)}{4\pi \xi^2 } \log \left(1 + \xi^2 \Lambda^2\right).
}
It is straightforward to apply formulae \eqref{fluc1},\eqref{fluc2} to obtain the modifications to the system-spanning fluctuations.

At the one-loop level, we could expand the functional determinant in powers of the sources to obtain the fluctuation corrections to the correlator; this is left for future work. Already at the classical level, there are  nontrivial results. Let us outline the computation of correlators in the saddle-point approximation. 

As discussed in the main text, we solve the classical equation \eqref{app_class} with sources 
\eq{
\alphab = \hat{\overline{\alpha}} + \sum_a \alphab^a \delta(\rv-\rv_a).
}
With $N$ sources we are able to compute $N$-body correlation functions. For simplicity, we will restrict consideration to $m=0$, as discussed in the main text. Then the saddle-point value of the action is
\eq{
S_c|_{m=0} = \int \left[ \alphab:\sigmab_c + \tilde\gamma \det \sigmab_c + 2 \tilde\eta p_c^2 - \nu \log p_c \right]
}
and we need
\eq{
\frac{\p^2 S_c}{\p \alpha^a_{ij} \p \alpha^b_{kl}} = \frac{\p \sigma_{ij}(\rv_a)}{\p \alpha^b_{kl}} + \frac{\p \sigma_{kl}(\rv_b)}{\p \alpha^a_{ij}} + & \int_r \left[ \alpha_{mn} \frac{\p^2 \sigma_{mn}}{\p \alpha^a_{ij} \p \alpha^b_{kl}} + \left(4 \tilde\eta + \frac{\nu}{p^2}\right) \frac{\p p}{\p \alpha^a_{ij}} \frac{\p p}{\p \alpha^b_{kl}} \right. \notag \\
& \qquad \left. + \left( 2 \tilde\eta p - \frac{\nu}{p} \right) \frac{\p^2 p}{\p \alpha^a_{ij} \p \alpha^b_{kl}}  + \tilde\gamma \check{\sigma}_{mn} \frac{\p^2 \sigma_{mn}}{\p \alpha^a_{ij} \p \alpha^b_{kl} } + \tilde\gamma \epsilon_{mp} \epsilon_{nq} \frac{\p \sigma_{mn}}{\p \alpha^a_{ij}  } \frac{\p \sigma_{pq}}{\p \alpha^b_{kl} }\right]
}
We evaluate the needed derivatives using the classical solution 
\eq{
 \psi_c(\rv) = \psi_s(\rv) + \frac{1}{4 \pi \tilde\eta}\int d^2 r' \; F\big[h(\rvp)-S(\rvp) \big] \; \log | \rv - \rvp|
}
where $F[h-S]= h - S + \sqrt{(h-S)^2 + 4 \tilde\eta \nu }$ and $S = (2\pi)^{-1} \sum_a \alphac^a : \nabla \nabla \log |\rv-\rv_a|$. The stress tensor is
\eq{ \label{app_nonlin4}
\sigmab_c(\rv) = \sigmabbar-\delb \; \pbar + \frac{1}{4\pi \tilde\eta}\int d^2 r' \; F\big[h(\rvp)-S(\rvp)\big]  \nabla \nabla \log |\rv-\rvp|
}
where 
\eq{
\nabla \nabla \log |\rv-\rvp| = \pi \delb \; \delta(\rv-\rvp) + \frac{1}{|\rv-\rvp|^2} \left[ \delb - 2 \nv \nv \right]
}
We need
\eq{
& \frac{\p \sigmab_c}{\p \alpha^a_{ij}  } = \frac{1}{4\pi \tilde\eta}\int d^2 r' \; F'_{r'} \; \nabla \nabla \log |\rv-\rvp| \frac{\p S_{r'}}{\p \alpha^a_{ij}} \\
& \frac{\p^2 \sigmab_c}{\p \alpha^a_{ij} \p \alpha^b_{kl} } = \frac{1}{4\pi \tilde\eta}\int d^2 r' \; F''_{r'} \; \nabla \nabla \log |\rv-\rvp| \frac{\p S_{r'}}{\p \alpha^a_{ij}} \frac{\p S_{r'}}{\p \alpha^b_{kl}} 
}

For the pressure-pressure correlator we need these expressions only when evaluated at $i=j$ and $k=l$.  We see that since $\p S_{r} / \p \alpha^a_{ii}  = \delta(\rv-\rv_a)$, most of the integrals drop out. It is not hard to see that
\eq{
\frac{\p^2 S_c}{\p \alpha^a_{ii} \p \alpha^b_{kk}} = A' \; \delta(\rv_a-\rv_b) + \frac{\gamma F'_{r_a}F'_{r_b}}{(4\pi \tilde\eta)^2} \int_{r} \epsilon_{mp} \epsilon_{nq} \big(\p_m \p_n \log |\rv-\rv_a| \big) \big( \p_p \p_q \log |\rv-\rv_b| \big)
}
for some $A'$ that collects all the contact terms. In an asymptotically large domain we can shift $\rv$ and rescale out $|\rv_a-\rv_b|$ to find
\eq{
\frac{\p^2 S_c}{\p \alpha^a_{ii} \p \alpha^b_{kk}} = 4 A \; \delta(\rv_a-\rv_b) + \frac{\gamma F'_{r_a}F'_{r_b}}{(4\pi \tilde\eta)^2} \frac{1}{|\rv_a-\rv_b|^2} I, 
}
where
\eq{
I & = \int_{r} \epsilon_{mp} \epsilon_{nq} \big(\p_m \p_n \log |\rv-\vec{e_x}| \big) \big( \p_p \p_q \log |\rv| \big) \\
& = -2 \int \frac{dr}{r} \int d\theta \; \frac{r^2 - 2 r \cos \theta + \cos 2\theta}{(r^2-2r \cos\theta+1)^2} \\
& = -2\pi \left(1 - \frac{|\rv_a-\rv_b|^2}{R^2} \right),
}
where we assumed that the shifted domain is $r<R/|\rv_a-\rv_b|$, and used properties of the Poisson kernel. $A$ is a modified coefficient that includes the contact term depending on $\gamma$. Note that this result is only valid for $|\rv_a-\rv_b| \ll R$, since in general we should use a domain $r < R$ in the original, non-shifted variables, and incorporate boundary conditions.

These results are for $m=0$. If we seek a perturbative solution $\psi_c = \psi_c^{(0)} + m  \psi_c^{(1)} + \ldots$, then we easily find
\eq{
\nabla^2 \psi_c^{(1)} = \frac{\nabla^2 \log \nabla^2 \psi_c^{(0)}}{\sqrt{(h-S)^2 + 4 \tilde\eta \nu}}
}
The main effect of this perturbation is to renormalize $F$ in the correlator:
\eq{
F \to F_R \equiv F + m \frac{\nabla^2 \log F}{\sqrt{(h-S)^2 + 4 \tilde\eta \nu}}
}


\end{widetext}

\end{document}